\documentclass{aa}  

\usepackage{graphicx}
\usepackage{txfonts}
\usepackage{siunitx,subfig,xspace,natbib}
\usepackage{hyperref}
\usepackage{amsmath}%
\usepackage{amsfonts}%
\usepackage{amssymb}%
\usepackage{mathrsfs}
\usepackage{arydshln}
\usepackage{cprotect}
\usepackage{multirow}

\usepackage{lipsum}

\newcommand{\mrm}[1]{\mathrm{#1}}

\newcommand{\nuc}[2]{$\mrm{^{#2}#1}$}

\begin{document}

   \title{Time-variable diffuse $\gamma$-ray foreground}


\author{
  Thomas Siegert
}
\institute{
	Julius-Maximilians-Universität Würzburg, Fakultät für Physik und Astronomie, Institut für Theoretische Physik und Astrophysik, Lehrstuhl für Astronomie, Emil-Fischer-Str. 31, D-97074 Würzburg, Germany
  	\label{inst:jmu}\\
  	\email{thomas.siegert@uni-wuerzburg.de}
}

   \date{Received XX, 2023; accepted MM DD, YYYY}

\abstract
{Observations of weak, serendipitous and possibly transient $\gamma$-ray signals often trigger large campaigns of follow-up studies.
	While the data analysis of $\gamma$-ray telescopes has now become more robust, these signals may just be misinterpretations of a time-variable foreground emission from the Solar System, induced by low-energy cosmic-ray interactions with asteroids.}
%
{Our goal is to provide emission templates for the time-variable diffuse $\gamma$-ray foreground by considering the populations of Main Belt Asteroids, Jovian and Neptunian Trojans, Trans Neptunian Objects (Kuiper Belt Objects), as well as the Oort Cloud.}
%
{By using the Small-Body Database, we obtain the spatial distribution of all known asteroids.
	We perform 3D-fits to determine their density profiles and calculate their appearances by line-of-sight integrations.
	Because Earth and the asteroids are moving with respect to each other, we obtain diffuse emission templates varying on timescales of days to decades.}
%
{We find that the temporal variability of the individual components can lead to flux enhancements which may mimic emission region unless properly taken into account in data analyses.
	Depending on photon energy, this variation is further enhanced by the Solar cycle as the cosmic-ray spectrum is attenuated by the Solar modulation potential, leading to a relative flux increase of the outer asteroid populations.
	The cumulative effect of the time-dependent emission is illustrated for the case of the 511\,keV `OSSE fountain', as well as for emission features near the Galactic Centre, both being possible misinterpretations of the Solar System albedo.}
%
{While the absolute luminosities for the asteroid populations are uncertain by one to three orders of magnitude, we recommend that $\gamma$-ray data analyses should always take into account the possibility of a time-variable foreground.
	Due to the ecliptic overlap with the Galactic plane, the Galactic emission is expected to be weaker by 0.1--20\%, depending on time (relative planetary motion), energy, and Solar cycle, which has immense consequences for the interpretation of dark matter annihilation cross sections, cosmic-ray spectra and amplitudes, as well as nucleosynthesis yields and related parameters.}


\maketitle
%

\section{Introduction}\label{sec:intro}
Space $\gamma$-ray telescopes in the MeV--GeV range measure photons indirectly by their interactions with the instruments' materials \citep{Siegert2022_gammaraytelescopes}.
This means that an interaction of two photons with identical energy and origin with the instrument will result in different appearances in the telescopes' native data spaces.
This process is generally termed dispersion and leads to the necessity of using pre-defined models to analyse the data.
While this effect is weaker in pair-creation telescopes, the dispersion in MeV instruments, be it coded aperture masks or Compton telescopes, is huge.
The large instrumental background in MeV telescopes makes it even more difficult to interpret the data without acknowledging the imaging response function.
A simple back-projection of recorded photons is therefore not appropriate and can lead to largely wrong interpretations.
Such misinterpretations may become even more severe if the diffuse emission that is measured by $\gamma$-ray telescopes appears to be variable in time.
Unless the temporal change of emission templates is properly taken into account, their cumulative effect `shines through' and potentially creates imaging artefacts or enhanced emission in places with no counterparts in other wavelengths, which in turn may create conjectures about additional emission mechanisms or exotic phenomena.
One such time-variable diffuse $\gamma$-ray foreground may originate in the cosmic-ray induced $\gamma$-ray albedo from asteroids in the Solar System \citep{Moskalenko2008_GRalbedoSS}.

\begin{figure*}[!t]%
	\centering
	\includegraphics[width=1.0\textwidth]{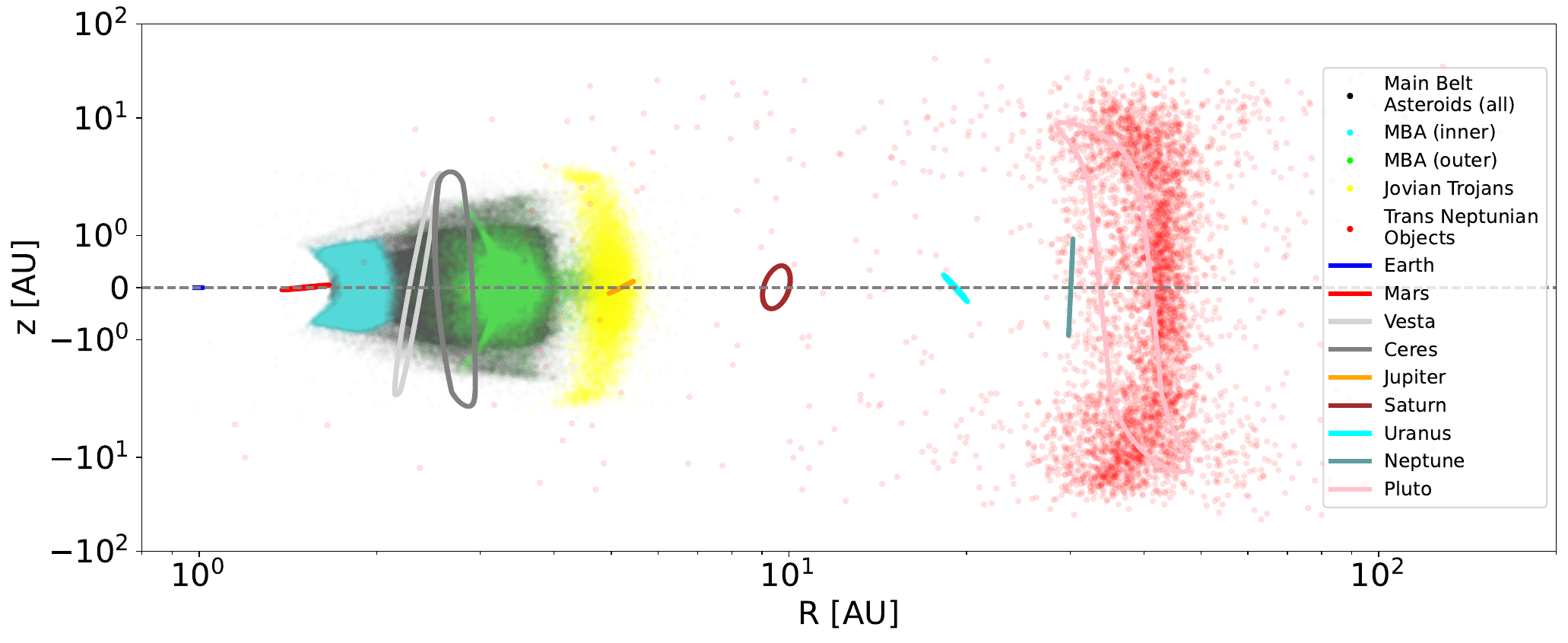}\\
	\includegraphics[width=0.3\textwidth]{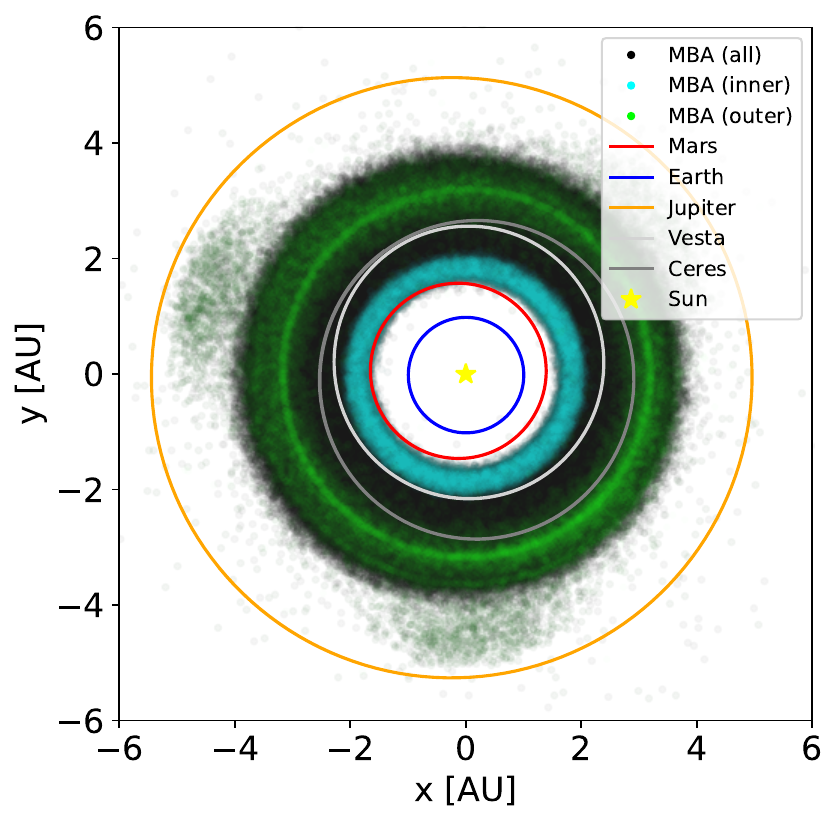}~
	\includegraphics[width=0.315\textwidth]{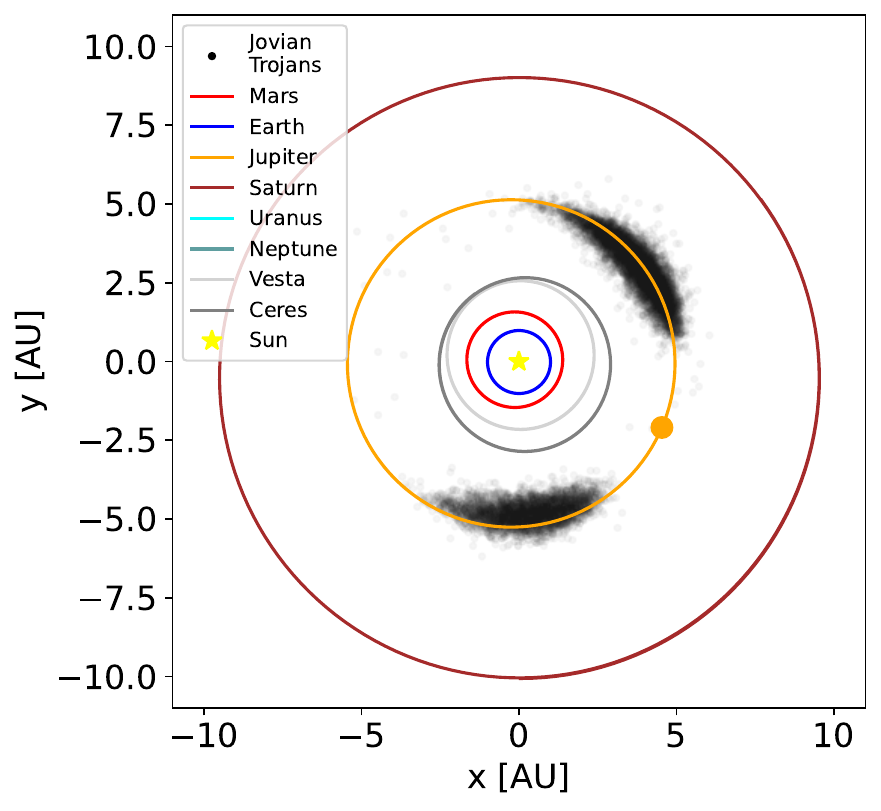}~
	\includegraphics[width=0.3125\textwidth]{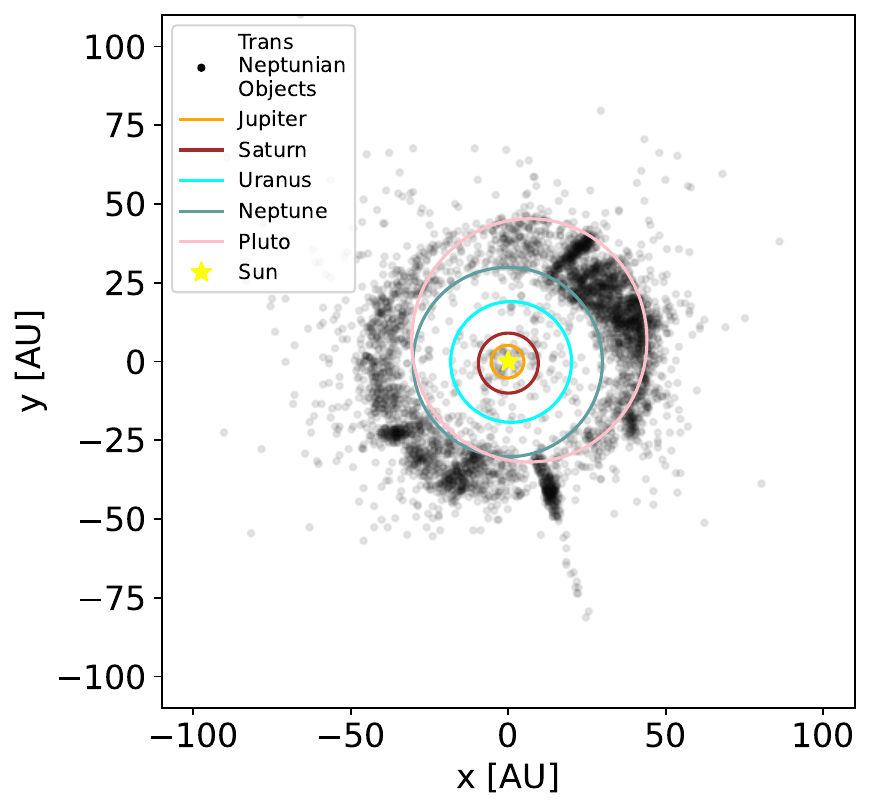}
	\caption{Spatial distribution of selected asteroids. \emph{Top}: Radial distribution of all asteroids in our study. \emph{Bottom}: Top view of asteroids, separated, from left to right, into Main Belt Asteroids, Jovian Trojans, and Trans Neptunian Objects.}
	\label{fig:all_asteroids}
\end{figure*}

\citet{Moskalenko2008_GRalbedoSS} calculated the $\gamma$-ray albedo spectrum of small Solar System objects with radial sizes above 100\,cm from 0.1\,MeV to 10\,GeV.
Their work is based on the $\gamma$-ray albedo of the Moon \citep{Moskalenko2007_GRalbedoMoon} which is known to shine in MeV--GeV photons, based on measurements from CGRO/EGRET \citep{Thompson1997_MoonSunEGRET} and then proven again by Fermi/LAT \citep{Loparco2017_FermiLAT_Moon}.
The Lunar Prospector mission \citep[LP;][]{Prettyman2006_MoonGammaRays} measured the albedo spectrum of the Moon in an orbit of only 100\,km above the lunar surface which resulted in maps of the Moon in different $\gamma$-ray lines according to its surface composition.
In particular, stable elements (O, Mg, Al, Si, Ca, Ti, and Fe, as well as composites) and radioactive elements (K, Th, and U) within a few tens of centimeters of the lunar surface have been measured with flux levels on the order of $10^{-3}$--$10^{-1}\,\mathrm{ph\,cm^{-2}\,s^{-1}}$ for a pixel size of $25\,\mathrm{deg^2}$ between 0.1 and 9.0\,MeV.
Given the distance to the Moon and the values from the LP, the calculated values from \citet{Moskalenko2007_GRalbedoMoon} are within a factor of 2--3 of the measured values.
However, the LP measurements focussed on $\gamma$-ray lines are did not discuss the expected bremsstrahlung spectrum, for example.
While \citet{Moskalenko2008_GRalbedoSS} discuss that there should be a time variability of the asteroid albedo, in this work, we explicitly calculate the appearance for any point in time for the different major asteroid accumulations:
the Main Belt Asteroids, the Jovian Trojans, the Neptunian Trojans, the Kuiper Belt Objects, and the unproven Oort Cloud.

The $\gamma$-ray albedo of individual asteroids has already been measured:
The Near-Earth Asteroid Rendezvous (NEAR) mission had an X- and $\gamma$-ray spectrometer on board its mission to the asteroid 433\,Eros \citep{Peplowski2016_Eros433}.
NEAR's closest approach to 433\,Eros was about 35\,km in which the 511\,keV electron-positron annihilation line as well as other elements (Fe, Mg, K, Si, Al, C, O, and composites) have been detected by their nuclear de-excitation above a bremsstrahlung spectrum.
Extrapolating the $\gamma$-ray spectrum towards the entire population of asteroids in all accumulations in the Solar System from this single $11.2 \times 11.2 \times 34.4\,\mathrm{km^3}$ asteroid is difficult if not meaningless as there are only a few other asteroids measured in soft $\gamma$-rays.
Ceres and Vesta, the largest asteroids in the Solar System with radii of $473$ and $263$\,km, respectively, have been visited by the Gamma Ray and Neutron Detector (GRaND) aboard the Dawn spacecraft \citep{Peplowski2013_Vesta_gamma,Lawrence2018_Ceres_gamma}.
Other objects without an atmosphere, such as Mercury \citep{Evans2012_Mercury_gamma} as well as the Mars moon Phobos \citep{Lawrence2019_Phobos_gamma}, have also been studied in terms of their surface composition, however not in terms of the bremsstrahlung induced by the same cosmic-ray spectrum at higher energies.
A Solar-System-wide analysis of these measurements would constrain the impact of the Sun's modulation potential in the inner Solar System as a function of distance even better than from measurements of the Pioneer and Voyager probes from the outer Solar System \citep{Fujii2005_CRheliosphere,Moskalenko2006_HeliosphericModulation}.
In any case, the points of reference are sparse for which reason we assume a generic asteroid $\gamma$-ray albedo spectrum at soft (MeV) and high-energy (GeV) $\gamma$-rays, which could be taken from \citet{Moskalenko2008_GRalbedoSS}, for example.
Since we are interested in the temporal variability and the appearance of the population of asteroids, the intrinsic spectrum is of less interest in this work.
In a future work, however, a more accurate model of the intrinsic cosmic-ray induced albedo spectrum of asteroids will help to disentangle the ecliptic from the Galactic emission.
In this context, \citet{Mesick2018_GEANT4_SSSB} showed that previous GEANT4 \citep{Agostinelli2003_GEANT4} versions, which were used to simulate the cosmic-ray impact on and $\gamma$-ray albedo from solid bodies, overpredict the measured values from the Apollo 17 Lunar Neutron Probe Experiment.

By measuring the time-variable diffuse $\gamma$-ray albedo from the population of asteroids in the Solar System, the Local Interstellar Cosmic-ray spectrum \citep{Vos2015_LICRS} could be constrained.
In addition would this foreground emission impact the Galactic diffuse emission \citep{Strong2005_gammaconti,Siegert2022_diffuseemission}, and in particular the shape and flux, and consequently the luminosity, of the Galactic bulge in GeV photons \citep[e.g.,][]{Ackermann2017_GeVGCE} or 511\,keV emission from electron positron annihilation \citep[e.g.,][]{Siegert2022_511,Siegert2023_511}.
The $\gamma$-ray albedo of the Solar System would also present a new possibility to detect so far elusive components, such as the Oort Cloud \citep{Hills1981_HillsCloud,Emelyanenko2007_OortCloud}, the population of Neptunian Trojans \citep{Sheppard2006_NeptunianTrojans}, and Kuiper Belt Objects (Trans Neptunian Objects) \citep{Schmedemann2017_SSSB}.
While individual objects are too faint to be observable with MeV--GeV instruments, whole asteroid populations are potentially detectable.
The spectral shape and absolute normalisation would then constrain the average composition of asteroids as well as their total number.

By properly taking into account also faint -- but spatially variable -- extended emission from the asteroids in the Solar System, apparent image artefacts and serendipitous other components could be minimised in image modelling and reconstructions.
The cumulative effect of moving asteroid accumulations with respect to a moving observer (Earth) will result in large-scale structures along the ecliptic with emission peaks at positions that depend on the observation period.
In particular the Jovian and Neptunian Trojans can potentially mimic emission near the Galactic bulge and disk which may then be falsely attributed to being Galactic phenomena rather than ecliptic ones.
In this work, we will work out the qualitative behaviour of the Solar System $\gamma$-ray albedo at MeV (soft) and GeV (high-energy) photon energies as it changes throughout a sidereal year as well as on longer timescales since the dawn of $\gamma$-ray observations in the 1960s.
Absolute numbers for fluxes of different accumulations are not available since the structures have not yet been detected as a whole \citep[see, however, flux estimates from][]{Moskalenko2008_GRalbedoSS};
this work intends to provide emission template maps for each point in time of previous and future $\gamma$-ray observations.

This paper is structured as follows:
In Sect.\,\ref{sec:sssb_definitions}, we describe small Solar System body accumulations and the database from which we select our objects.
Sect.\,\ref{sec:modelling} contains the details of modelling 3D-density distributions of Solar System asteroids.
We describe the line-of-sight effects of different accumulations with respect to a moving observer (Earth) in Sect.\,\ref{sec:los}.
Contentious or special emission feature, especially in the soft $\gamma$-ray band but also at high-energy $\gamma$-rays, are evaluated in the context of ecliptic emission from asteroids in Sect.\,\ref{sec:image_artefacts}.
We discuss our findings considering the detectability with current and future instruments in the $\gamma$-ray regime, considering the absolute flux levels of Galactic emission when the albedo is taken into account, and recommendations for (future) works in Sect.\,\ref{sec:discussion}.
We conclude in Sect.\,\ref{sec:conclusion}.

\section{Small Solar system body accumulations}\label{sec:sssb_definitions}
Single asteroids are known to shine in $\gamma$-rays \citep{Peplowski2016_Eros433}, but are too faint to be observable individually with current instruments from large distances ($\gtrsim 1000$\,km).
Therefore, we used the Jet Propulsion Laboratory's (JPL's) Small-Body Database Query (SBDB)\footnote{\url{https://ssd.jpl.nasa.gov/tools/sbdb_query.html}} to describe the (known) population of asteroids accumulating in different parts of the Solar System.
For this study, we select by orbit classes, either the Main-Belt Asteroids, separated into Inner, Outer, and general, Jupiter Trojans, and Trans Neptunian Objects.
Other populations, for example Near Earth Asteroids or the Centaurs, are too sparse for the purpose of this work to create density distributions (Sect.\,\ref{sec:modelling}) for line-of-sight integrated emissivity profiles (Sect.\,\ref{sec:los_integral}).

In SBDB, we select the output fields \texttt{epoch\_mjd}, \texttt{e}, \texttt{a}, \texttt{i}, \texttt{node}, \texttt{peri}, and \texttt{M}, which include the epoch of osculation in modified Julian days (MJD; $T_o$), the eccentricity $e$, the semimajor axis $a$ in AU, the inclination $i$ with respect to the ecliptic plane ($xy$-plane) in degrees, the longitude of the ascending node $\Omega$ in degrees, the argument of perihelion $\omega$ in degrees, and the mean anomaly $M$ in degrees, respectively.
Given the ever-increasing number of detected asteroids of currently 6--7\,\% additional detections in the Main Belt per year, we freeze the time of our analysis and database downloads to 2023-07-13.
The total number of Main Belt Asteroids is then 1,223,983, with 27,877 Inner asteroids, and 39,868 Outer asteroids.
For the Jovian Trojans, we get 12,636 database entries, currently increasing at a rate of 4--5\,\% per year.
There are 4,464 Trans Neptunian Objects in the database, whose numbers also increase by 6--7\,\% per year.

While the number of detected Neptunian Trojans\footnote{\url{https://minorplanetcenter.net//iau/lists/NeptuneTrojans.html}} is currently below 50, the total number of Neptunian Trojans could exceed the number of Jovian Trojans by a factor of a few \citep{Sheppard2006_NeptunianTrojans}.
We will model the Neptunian Trojans by using the Jupiter Trojans as surrogate population and scaling them according to literature values (Sect.\,\ref{sec:neptunian_trojans}).
Likewise, the Trans Neptunian Object population beyond the Kuiper Belt may form another disk-like entity called the Hills cloud \citep{Hills1981_HillsCloud,Bailey1988_HillsCloud} up to $\sim 5000$\,AU, and a spherical accumulation known as the Oort Cloud, supposedly located outside the heliosphere between 2000\,AU and 50000--200000\,AU \citep[][and references therein]{Emelyanenko2007_OortCloud}.
While the existence of these accumulations is questionable, we will briefly discuss their impact on the $\gamma$-ray foreground in Sect.\,\ref{sec:oort_cloud}.

We illustrate the spatial distribution of our selection for the Main Belt Asteroids, Jovian Trojans, and Trans Neptunian Objects, respectively, in Fig.\,\ref{fig:all_asteroids} radially and in the ecliptic plane.
For reference, we show the major and largest minor bodies from Earth to Pluto.

\section{3D-Modelling of Solar system asteroid accumulations}\label{sec:modelling}
We use the catalogues from Sect.\,\ref{sec:sssb_definitions} and create 3D histograms of objects per unit volume.
These histograms are then fitted with specific 3D-density functions which we describe below in detail.
Depending on how many asteroids have already been detected for each subgroup, we change the binning accordingly to have a visual impression of how accurate our density functions fit.
We note that the binning has no influence on the fit parameters because we use the Poisson likelihood for counting a number of objects within a unit bin size as this is number conserving.
The 3D-densities we use have only a weak physical interpretation and serve as descriptive functions with fixed shape parameters and a free amplitude or normalisation.

We use the orbital elements to calculate the true anomaly $\nu$ from the eccentricity $e$ and the mean anomaly $M$.
Then, we calculate the Cartesian coordinates of each object in a heliocentric coordinate system at the observation / detection epoch $T_o$, and propagate them to a reference time $T_r = 59539$\,MJD (2021-11-21 00:00:00.000).
In this way, we can obtain the density function of the population of objects as a whole at any given time in the near past (several 100\,yr) and future (up to 100\,yr).
The Cartesian coordinates are binned into 3D histograms for each subgroup.

We use \texttt{Migrad} from the \texttt{Minuit2} library \citep{James1975_Minuit} with a Poissonian likelihood,
\begin{equation}
	\mathscr{L}(D|M(\boldsymbol{\phi})) = 2\sum_{x,y,z} \left[M_{xyz}(\boldsymbol{\phi}) - D_{xyz} \ln\left(M_{xyz}(\boldsymbol{\phi})\right)\right]\mrm{,}
	\label{eq:poisson_loglik}
\end{equation}
to fit our data $D$, which are counts in $xyz$-bins, with a model $M$ that depends on a set of parameters $\boldsymbol{\phi}$.
The uncertainties obtained in this way must not be over-interpreted as the detected asteroids are only a small sample of the actual number.

\subsection{Main Belt Asteroids}\label{sec:Main Belt Asteroids}
We model the torus-like asteroid accumulations, such as the Main Belt Asteroids or Trans Neptunian Objects, with sums of Gaussian tori,
\begin{equation}
	\rho_{\rm GT}(x,y,z;\rho_0,R,\Sigma,r,\sigma) = \rho_0 \exp\left[-\frac{1}{2}\left(\frac{\left(R_{xy} - R\right)^2}{\Sigma^2} + \frac{\left(z-r\right)^2}{\sigma^2}\right)\right]dV\mrm{,}
	\label{eq:Gaussian_torus}
\end{equation}
where $R_{xy}^2 = x^2 + y^2$, $R$ and $\Sigma$ are the large torus radius in the $xy$-plane and the radial width, $r$ and $\sigma$ are the vertical radial offset and vertical width, and $dV$ is the volume element, respectively.
The amplitude $\rho_0$ serves as a normalisation constant in units of number of asteroids per $\mrm{AU}^3$, and scales the different tori with respect to each other.
The integral of Eq\,(\ref{eq:Gaussian_torus}), $I(R,\Sigma,\sigma)$, is analytically solvable 
so that the total number of objects included in the fit can be re-obtained by $N_{\rm obj} = I(R,\Sigma,\sigma) \times \rho_0$.
The emissivity of the accumulations will be discussed in Sect.\,\ref{sec:los}.

\begin{table}[t]
	\begin{center}
		\centering
		\caption{Fit parameters for Gaussian tori to describe the asteroid number densities, Eq.\,(\ref{eq:Gaussian_torus}), for Main Belt Asteroids and subgroups. The total Main Belt is described by the sum of five, 1--5, tori. The units are unitless for the number of objects $N_{\rm obj}$, $10^3\,\mathrm{AU^{-3}}$ for $\rho_0$, and $\mathrm{AU}$ for the other parameters.}\label{tab:Main Belt Asteroid_fit_parameters}%
		\begin{tabular}{l || r |rrrrr}
			\hline
			
			Group & $N_{\rm obj}$ & $\rho_0$ & $R$ & $\Sigma$ & $r$ & $\sigma$ \\
			
			\hline
			Inner & $27877$ & $6.15$ & $1.845$ & $0.125$ & $0.005$ & $0.502$ \\
			Outer & $39868$ & $1.27$ & $3.725$ & $0.450$ & $0.003$ & $0.542$ \\
			Main 1 & $409370$ & $89.53$ & $2.333$ & $0.312$ & $-0.002$ & $0.159$ \\
			Main 2 & $703556$ & $45.24$ & $2.814$ & $0.358$ & $0.006$ & $0.391$ \\
			Main 3 & $88353$ & $1.76$ & $3.043$ & $0.578$ & $-0.012$ & $0.723$ \\
			Main 4 & $11566$ & $7.67$ & $1.840$ & $0.137$ & $-0.559$ & $0.151$ \\
			Main 5 & $11138$ & $7.64$ & $1.826$ & $0.132$ & $0.575$ & $0.153$ \\
			\hline
		\end{tabular}
	\end{center}
\end{table}

\begin{figure*}[!t]%
	\centering
	\includegraphics[width=1.0\textwidth]{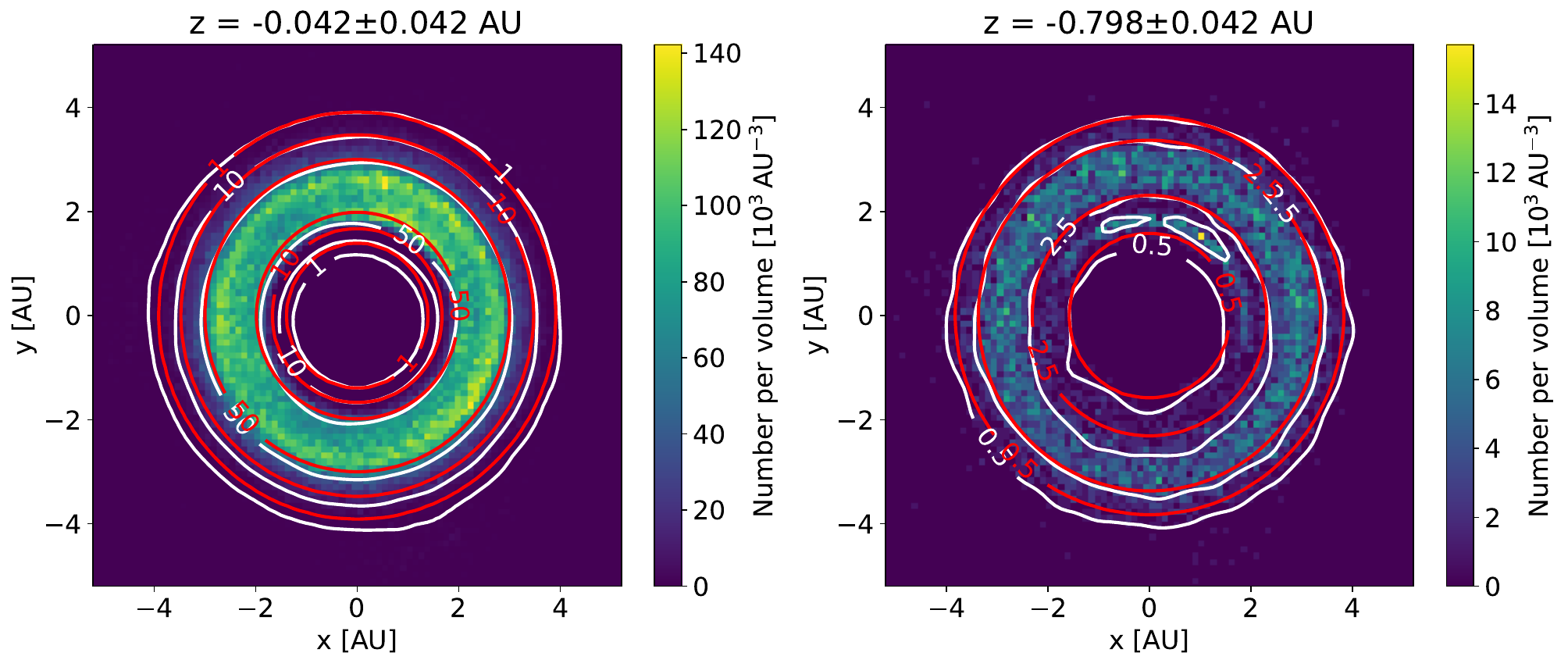}\\
	\includegraphics[width=1.0\textwidth]{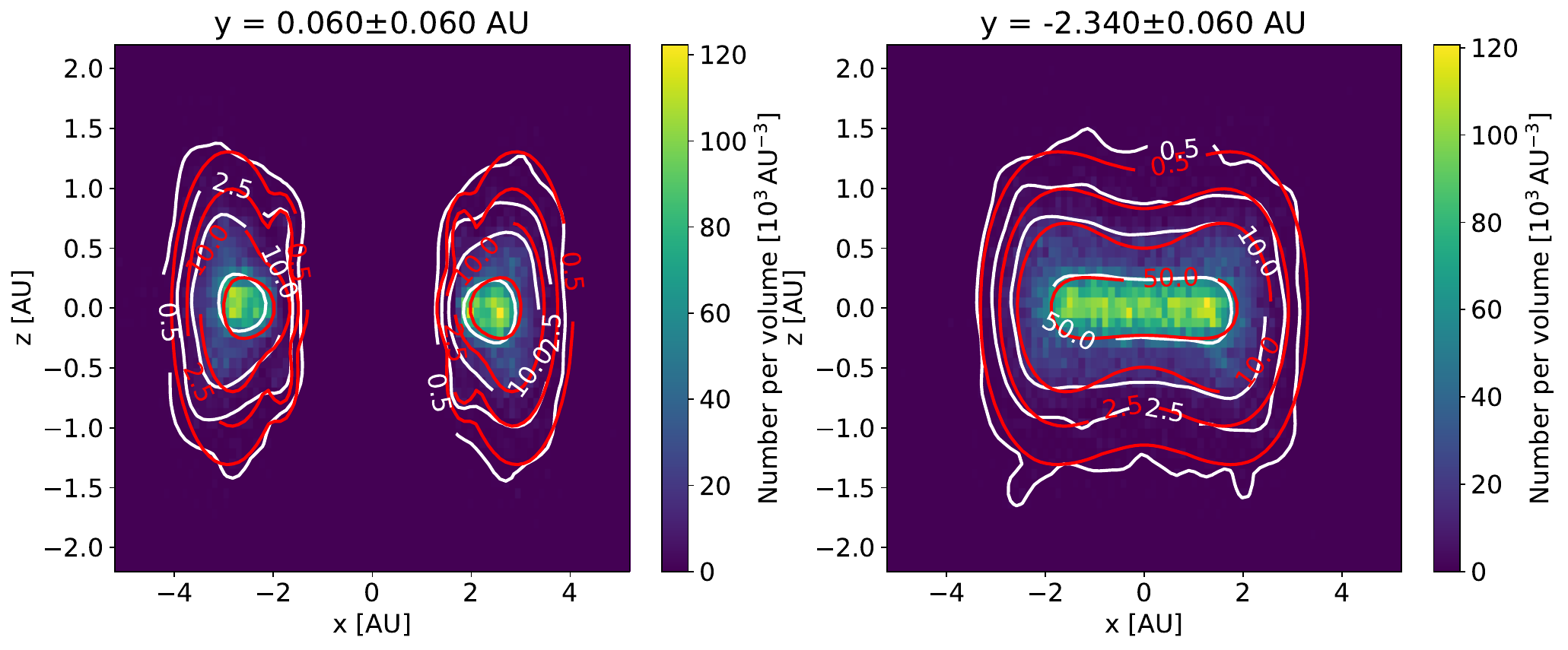}
	\caption{Distribution of Main Belt Asteroids binned into 3D-pixels. Shown are slices (coloured 2D-histograms) of the $xy$-plane at different $z$-values (top) and of the $xz$-plane at different $y$-values (bottom). The white contours show the smoothed distributions of the 2D-histograms in each panel. The fitted models are shown as red contours, capturing most of the important, that is high-density, features, as well as the major wings.}
	\label{fig:Main Belt Asteroid_density}
\end{figure*}

\subsubsection{Hungaria Asteroids}\label{sec:Main Belt Asteroid_inner}
The Hungaria family of asteroids, also called Inner Main Belt Asteroid, are found roughly between 1.6 and 2.1\,AU as part of the Main Belt Asteroids.
Their eccentricities range between $0$ and $0.16$, and their inclinations peak around $22^\circ$.
Given the distance and inclination distribution, we can expect the radius, width, and vertical width to be around 1.8\,AU, 0.125\,AU, and 0.35\,AU, respectively. 
For the fits to converge in a reasonable time, we will always use these characteristic values as starting points.
We fit one Gaussian torus to the Inner Main Belt Asteroids, which we bin in $80 \times 80 \times 60 = 384,000$ bins in $x$, $y$, and $z$, respectively, ranging in the intervals $\left[x_{\rm min},x_{\rm max}\right] = \left[-2.5,2.5\right]$\,AU, $\left[y_{\rm min},y_{\rm max}\right] = \left[-2.5,2.5\right]$\,AU, and $\left[z_{\rm min},z_{\rm max}\right] = \left[-1.5,1.5\right]$\,AU.
Thus, the 3D-bin size is $\Delta V = \Delta x \times \Delta y \times \Delta z = 0.0625\,\mathrm{AU} \times 0.0625\,\mathrm{AU} \times 0.05\,\mathrm{AU} = 1.953125 \times 10^{-4}\,\mathrm{AU^3}$.
The fitted parameters are summarised together with the other Main Belt Asteroid accumulations in Tab.\,\ref{tab:Main Belt Asteroid_fit_parameters}.
We restrict ourselves and only illustrate the model fits of the complete Main Belt Asteroids, Jovian Trojans, and Trans Neptunian Objects in different slices.

\subsubsection{Outer Asteroids}\label{sec:Main Belt Asteroid_outer}
The Outer asteroids of the Main Belt are found between 2 and 5\,AU with large accumulations around $3.2 \pm 0.5$\,AU, with eccentricities from $0$ to $0.7$, mostly below $0.35$, and inclinations peaking around $10^\circ$ with values up to $40^\circ$.
The Hilda asteroids are also part of the Outer Main Belt, making a triangle-like structure in the ecliptic plane between Mars and Jupiter.
They are only a small fraction of the Outer asteroids, so that we will omit them in the modelling.
We use $100 \times 100 \times 80 = 800,000$ $xyz$-bins in the intervals $\left[-7.0,7.0\right]$\,AU, $\left[-7.0,7.0\right]$\,AU, and $\left[-4.2,4.2\right]$\,AU, so that the 3D-bin size is $\Delta V = \Delta x \times \Delta y \times \Delta z = 0.14\,\mathrm{AU} \times 0.14\,\mathrm{AU} \times 0.105\,\mathrm{AU} = 2.058 \times 10^{-3}\,\mathrm{AU^3}$.
We again find that one Gaussian torus is enough to describe the subgroup, and the fit parameters are shown in Tab.\,\ref{tab:Main Belt Asteroid_fit_parameters}.

\subsubsection{Complete Belt}\label{sec:Main Belt Asteroid_complete}
For the Main Belt Asteroids as a whole, we use five Gaussian tori that describe the accumulations in the full range up to 7.2\,AU radially with the peak around 2.7\,AU, eccentricities peaking around $0.14$, and inclinations around $7^\circ$, with large wings in the distributions due to the Inner and Outer Main Belt Asteroids.
We bin the more than one million objects in $120 \times 120 \times 100 = 1,440,000$ $xyz$-bins in the intervals $\left[-7.2,7.2\right]$\,AU, $\left[-7.2,7.2\right]$\,AU, and $\left[-4.2,4.2\right]$\,AU, for a 3D-bin size of $\Delta V = \Delta x \times \Delta y \times \Delta z = 0.12\,\mathrm{AU} \times 0.12\,\mathrm{AU} \times 0.084\,\mathrm{AU} = 1.2096 \times 10^{-3}\,\mathrm{AU^3}$.
The fitted parameters are shown in Tab.\,\ref{tab:Main Belt Asteroid_fit_parameters}.
We note that the total number of objects for the sum of Gaussian tori is given by $\sum_{i=1}^{5} I_i(R_i,\Sigma_i,\sigma_i) \times \rho_{0,i}$, from which the number of objects contributing to each torus can be estimated (second column in Tab.\,\ref{tab:Main Belt Asteroid_fit_parameters}).

As an example of how the density distributions appear when binned, we show the total accumulation of Main Belt Asteroids in Fig.\,\ref{fig:Main Belt Asteroid_density} together with the fitted model.
It becomes evident that the larger structure is well represented by the combination of five Gaussian tori, and even peculiarities at the inner edge of the distribution (Fig.\,\ref{fig:Main Belt Asteroid_density}, bottom left) are captured.
Further modelling of a possible tilt in the $xy$-plane of the distribution and matching every subgroup by itself is beyond the purpose of this paper and may be left for a future study when such an accuracy is actually required.

\begin{figure*}[!t]%
	\centering
	\includegraphics[width=1.0\textwidth]{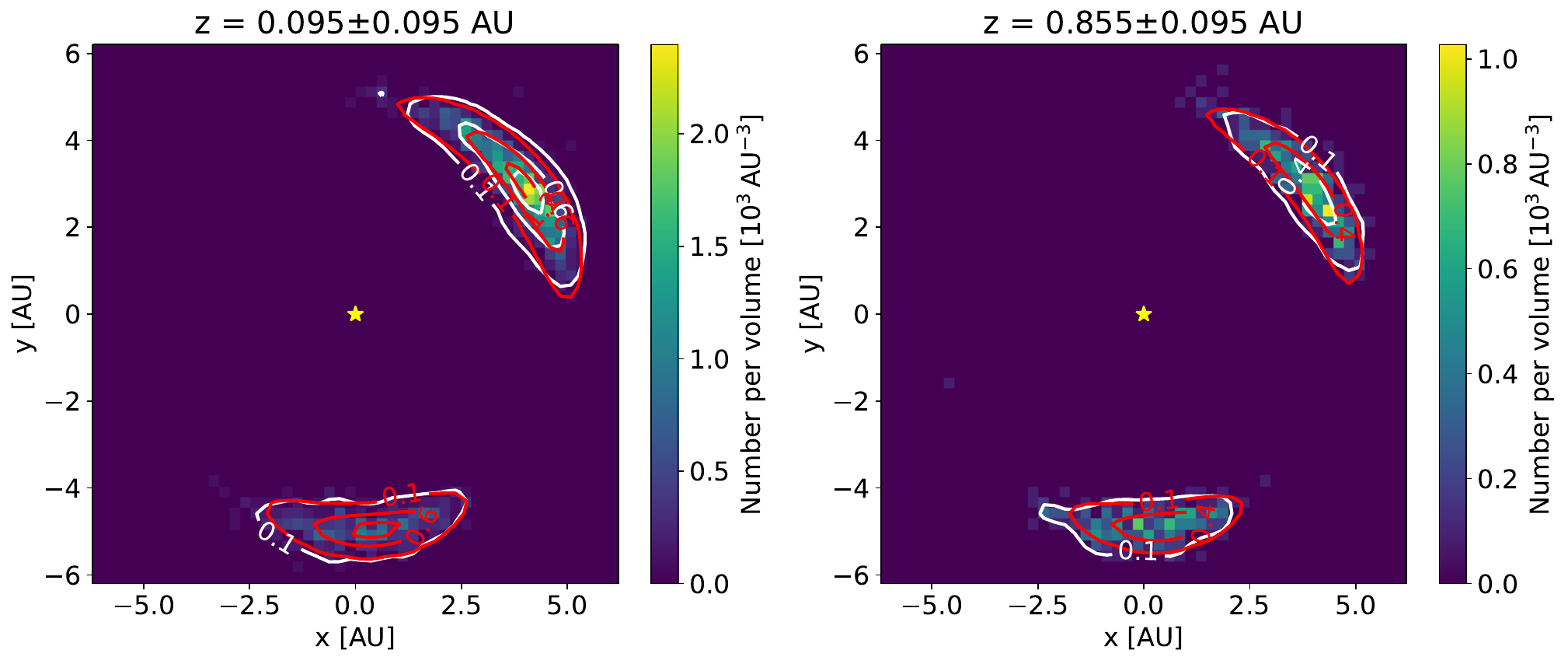}\\
	\includegraphics[width=1.0\textwidth]{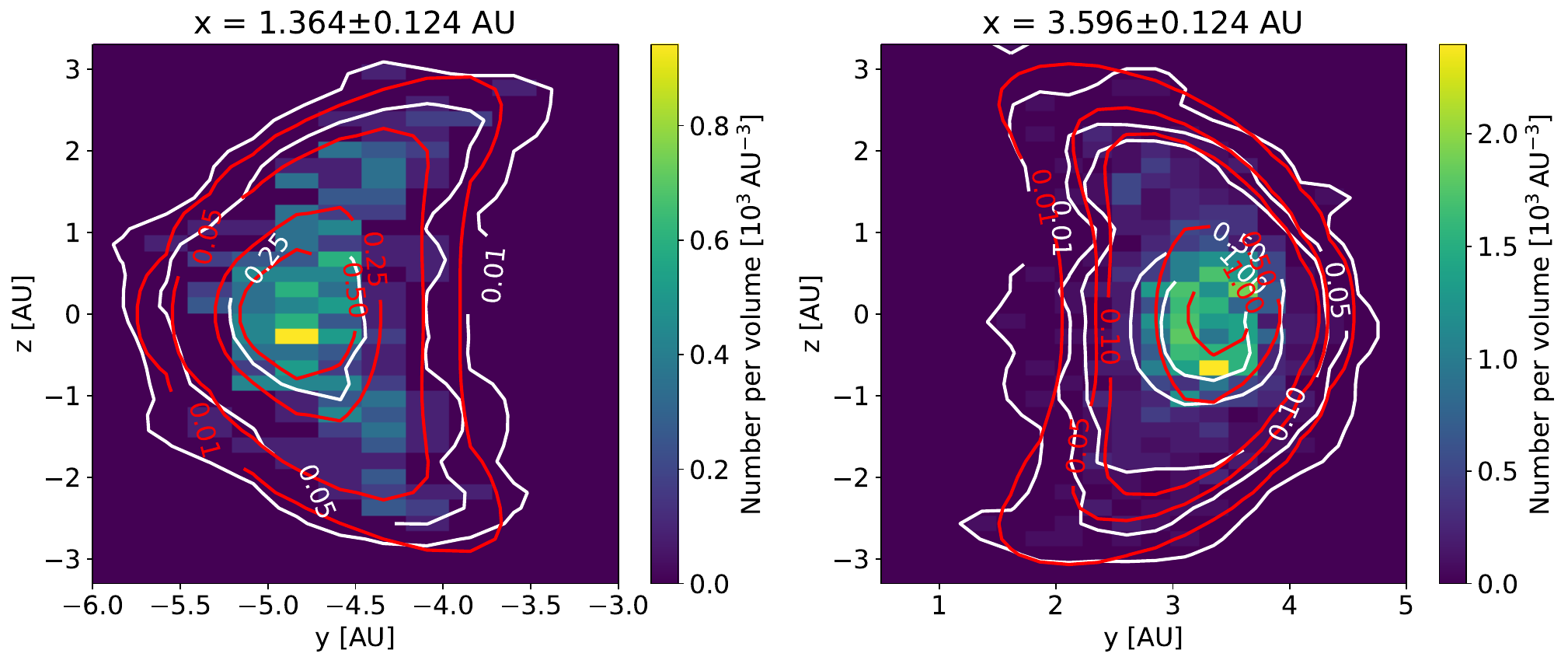}
	\caption{Same as Fig.\,\ref{fig:Main Belt Asteroid_density} but for the distribution of Jovian Trojans. For reference, the position of the Sun is shown in the top panel with the yellow star symbol. The bottom panel shows a zoom-in of the $yz$-plane to show the L4 and L5 trojans, respectively.}
	\label{fig:JT_density}
\end{figure*}

\subsection{Jovian Trojans}\label{sec:jovian_trojans}
The detected Jovian Trojans are distributed along the orbit of Jupiter so that their radial distribution also peaks at the orbital distance of Jupiter around 5\,AU, formally ranging between 3.7 and 6.5\,AU.
The trojans' eccentricities range between $0$ and $0.3$, with most of the objects to be found between $0$ and $0.15$.
The inclinations peak around $12^\circ$, and range up to $40^\circ$, so that the vertical width is expected to be 1\,AU or more.

We mimic the shape of the Jovian Trojans by a spherical Gaussian shell, subtended by two Gaussian ellipses at the Lagrange points L4 and L5, and additionally subtended by a Gaussian in $z$-direction centred in the ecliptic.
The Gaussian shell is described by
\begin{equation}
	\rho_{\rm GS}(x,y,z;\rho_s,R_s,\Sigma_s) = \rho_s\exp\left[-\frac{1}{2}\left(\frac{R_{xyz} - R_s}{\Sigma_s}\right)^2\right]dV\,\mrm{,}
	\label{eq:Gaussian_shell}
\end{equation}
where $R_{xyz}^2 = x^2 + y^2 + z^2$, $R_s$ is the radius of the shell, $\Sigma_s$ its width, $\rho_s$ the normalisation (in units of objects per $\mathrm{AU^3}$), and $dV$ the volume element.
A tilted Gaussian ellipse is described by
\begin{eqnarray}
	\rho_{\rm GE}(x,y,z;\rho_e,x_e,y_e,\sigma_x,\sigma_y,\theta) & = \nonumber\\
	\rho_e \exp\left[-\left(a(x-x_e)^2  + 2b(x-x_e)(y-y_e) + c(y-y_e)^2 \right)\right]\,\mrm{,}
	\label{eq:Gaussian_ellipse}
\end{eqnarray}
with
\begin{eqnarray}
	a & = & \frac{1}{2}\left[ \left(\frac{\cos\theta}{\sigma_x}\right)^2 + \left(\frac{\sin\theta}{\sigma_y}\right)^2 \right]\,\mrm{,}\\
	b & = & \frac{1}{4}\sin(2\theta)\left(-\frac{1}{\sigma_x^2} + \frac{1}{\sigma_y^2}\right)\,\mrm{,\,and}\\
	c & = & \frac{1}{2}\left[ \left(\frac{\sin\theta}{\sigma_x}\right)^2 + \left(\frac{\cos\theta}{\sigma_y}\right)^2 \right]\mrm{.}
\end{eqnarray}
In Eq.\,(\ref{eq:Gaussian_ellipse}), $\rho_e$ is again the normalisation, $x_e$ and $y_e$ are the coordinates of the ellipse, $\sigma_x$ and $\sigma_y$ its widths, and $\theta$ the tilt angle.
Finally, a vertical Gaussian centred in the ecliptic plane ($xy$-plane) is given by
\begin{equation}
	\rho_{\rm Gz}(x,y,z;\rho_z,\sigma_z) = \rho_z \exp\left[-\frac{1}{2}\left(\frac{z}{\sigma_z}\right)^2\right]\,\mrm{,}
	\label{eq:Gaussian_z}
\end{equation}
with $\rho_z$ being the normalisation and $\sigma_z$ the vertical width.
The functions $\rho_{\rm Gz} \times \rho_{\rm GE}$ describes a Gaussian ellipsoid, centred and tilted in the $xy$-plane.
The positions of the Lagrange points are fixed at the reference time $T_r$, and are calculated by knowing the Cartesian coordinates of Jupiter at this time ($x_j = 4.541$\,AU, $y_j = -2.094$\,AU, $z_j = -0.093$\,AU), which gives a distance to the Sun of $d_j = 5.002$\,AU, and the masses of Jupiter ($M_j = 317.8\,\mrm{M_\oplus}$) and the Sun ($M_\odot = 332950\,\mrm{M_\oplus}$).
Ignoring the small deviations from the $z$-component, we get the general Lagrange points from weighting with the reduced mass $\mu_{\odot,j} = \frac{M_\odot - M_j}{M_\odot + M_j}$ so that
\begin{eqnarray}
	x_{\rm L4,0} & = & \frac{1}{2} d_j \mu_{\odot,j} \,\mrm{and}\\
	y_{\rm L4,0} & = & \frac{\sqrt{3}}{2} d_j \mu_{\odot,j}
\end{eqnarray}
in the unrotated frame (general Lagrange points), which is converted to the rotated frame (current time) by a simple rotation in the $xy$-plane,
\begin{eqnarray}
	x_{\rm L4,j} & = & x_j \cos(\alpha) + y_j \sin(\alpha)\mrm{,}\\
	y_{\rm L4,j} & = & - x_j \sin(\alpha) + y_j \cos(\alpha)\mrm{,}\\
	x_{\rm L5,j} & = & x_j \cos(-\alpha) + y_j \sin(-\alpha)\mrm{,}\\
	y_{\rm L4,j} & = & - x_j \sin(-\alpha) + y_j \cos(-\alpha)\mrm{,}
	\label{eq:Lagrange_points}
\end{eqnarray}
where $\alpha = \arctan(y_{\rm L4,0},x_{\rm L4,0})$.
%
%

\begin{table*}[t]
	\begin{center}
		\centering
		\caption{Fit parameters for Eq.\,(\ref{eq:Jovian_trojans}) to describe the asteroid number densities of the Jovian and Neptunian Trojans. The units are $10^3\,\mathrm{AU^{-3}}$ for the densities $\rho_{\rm L4/5}$, $\mathrm{AU}$ for the scaling parameters, and $\mathrm{rad}$ for the angles $\theta_{\rm L4/5}$.}\label{tab:JT_fit_parameters}%
		\begin{tabular}{l || l | rrrrrrrrrr}
			\hline
			
			Planet & Parameter & $\rho_{\rm L4}$ & $\rho_{\rm L5}$ & $R_s$ & $\Sigma_s$ & $\sigma_{x}$ & $\sigma_{y}$ & $\sigma_{\rm z,L4}$ & $\sigma_{\rm z,L5}$ & $\theta_{\rm L4}$ & $\theta_{\rm L5}$ \\
			
			\hline
			
			Jupiter & Value & $0.79$ & $0.70$ & $4.989$ & $0.278$ & $0.974$ & $1.613$ & $1.120$ & $0.488$ & $5.115$ & $-0.487$ \\
			Neptune & Value & $0.79$ & $0.79$ & $28.808$ & $1.605$ & $3.975$ & $6.585$ & $5.849$ & $2.550$ & $5.115$ & $-0.487$ \\
			
			\hline
		\end{tabular}
	\end{center}
\end{table*}

Finally, the function we use to fit the asteroid distribution of the Jovian Trojans reads
\begin{eqnarray}
	\rho_{\rm JT}(x,y,z;\rho_{\rm L4},\rho_{\rm L5},R_s,\Sigma_s,\sigma_{x},\sigma_{y},\sigma_{\rm z,L4},\sigma_{\rm z,L5},\theta_{\rm L4},\theta_{\rm L5}) & = & \nonumber\\
	\rho_{\rm GS}(x,y,z;1,R_s,\Sigma_s) & \times & \nonumber\\
	\left[\rho_{\rm Gz}(x,y,z;\rho_{\rm L4},\sigma_{\rm z,L4}) + \rho_{\rm Gz}(x,y,z;\rho_{\rm L5},\sigma_{\rm z,L4})\right] & \times & \nonumber\\
	\left[\rho_{\rm GE}(x,y,z;1,x_{\rm L4,j},y_{\rm L4,j},\sigma_x,\sigma_y,\theta_{\rm L4})\right. & + & \nonumber\\
	\left.\rho_{\rm GE}(x,y,z;1,x_{\rm L5,j},y_{\rm L5,j},\sigma_x,\sigma_y,\theta_{\rm L5})\right]dV\mrm{.}
	 \label{eq:Jovian_trojans}
\end{eqnarray}
The ten free parameters of Eq.\,(\ref{eq:Jovian_trojans}) take into account that the distribution of Jovian Trojans is asymmetric, that is, there are about twice as many asteroids in L4 than in L5.
While this may be a detection bias, we allow the model to obtain different normalisations $\rho_{\rm L4}$ and $\rho_{\rm L5}$ for the two Lagrange points, respectively.
We fix the centroid of the trojan accumulations to the coordinates of the Lagrange points, $x_{\rm L4/5,j}$ and $y_{\rm L4/5,j}$, which have an explicit dependence on time as Jupiter and its trojans are propagating around the Sun.
This will lead to an apparently moving foreground emission, further discussed in Sect.\,\ref{sec:los}.
The radial widths of the Jovian Trojan asteroid populations, $\sigma_{x}$ and $\sigma_{y}$ in combination with $\Sigma_s$, are assumed to be the same for L4 and L5, however the vertical extents, $\sigma_{\rm z,L4}$ and $\sigma_{\rm z,L5}$ may change independently.
The distribution is always bound to the large spherical shell radius $R_s$.

We bin the Jovian Trojan data into $50 \times 50 \times 40 = 100,000$ $xyz$-bins in the intervals $\left[-6.2,6.2\right]$\,AU, $\left[-6.2,6.2\right]$\,AU, and $\left[-3.8,3.8\right]$\,AU, for a 3D-bin size of $\Delta V = \Delta x \times \Delta y \times \Delta z = 0.248\,\mathrm{AU} \times 0.248\,\mathrm{AU} \times 0.19\,\mathrm{AU} = 1.168576 \times 10^{-2}\,\mathrm{AU^3}$.
The fitted parameters are shown in Tab.\,\ref{tab:JT_fit_parameters}, and the comparison of the binned data with the fitted model in Fig.\,\ref{fig:JT_density}.
It is clear that our model describes the asymmetry of the asteroid number in L4 and L5 properly, both radially and vertically.
We note that the asteroid number densities in L4 and L5 are, in fact, not too different, but the vertical extent is 2--3 times as large in L4 than in L5 which captures more of the asteroids at higher $z$, that is, higher inclinations.

\subsection{Neptunian Trojans}\label{sec:neptunian_trojans}
Because only $\sim 30$ Neptunian Trojans are detected so far, but many more expected, we use the model of the Jovian Trojans from Sect.\,\ref{sec:jovian_trojans} and scale it in accordance to the planets' parameters:
The mean heliocentric distance of Neptune ($d_n = 30.047$\,AU) compared to Jupiter ($d_j = 5.204$\,AU) is used to scale the radial parameters of our model to describe the trojans, Eq.\,(\ref{eq:Jovian_trojans}), by multiplying with the distance ratio $r_{nj} = d_n/d_j = 5.774$.
Thus, the parameters of the spherical shell, $R_s$ and $\Sigma_s$ will change ($R_s \rightarrow r_{nj} R_s$; $\Sigma_s \rightarrow r_{nj} \Sigma_s$), as well as the $x$- and $y$-widths of the Gaussian ellipses ($\sigma_x \rightarrow \frac{r_{nj}}{\sqrt{2}}\sigma_x$; $\sigma_y \rightarrow \frac{r_{nj}}{\sqrt{2}}\sigma_y$).
For the vertical extent, we consider the median inclinations of the discovered Neptune Trojans ($i_n = 18^\circ$) and of the Jovian Trojans ($i_j = 12^\circ$) and calculate a scaling according to the tangents of the median inclinations, $\iota_{nj} = \tan(i_n)/\tan(i_j) = 1.567$, and also considering the widening vertically taking into account the radial stretch, so that $\xi_{nj} = \frac{\iota_{nj}r_{nj}}{\sqrt{3}} = 5.223$.
We use this value to adapt the vertical widths $\sigma_{\rm z,L4} \rightarrow \xi_{nj}\sigma_{\rm z,L4}$ and $\sigma_{\rm z,L5} \rightarrow \xi_{nj}\sigma_{\rm z,L5}$.
While this vertical scaling may be wrong by a factor of a few, it will provide a first-order estimate of the possible distribution of a $\gamma$-ray albedo from the Neptunian Trojans.
Because the orbital period of Neptune is $T_n = 164.8$\,yr and its mass is also smaller than that of Jupiter ($M_n = 0.01715\,\mathrm{M_\oplus}$), the Lagrange points will slightly change and move slowlier across the sky.
This is taken into account by calculating the distribution as it propagates from the reference time to any specified time.
Finally, we only consider a symmetric distribution of Neptunian Trojans in L4 and L5, even though also the discovered trojans are much more frequent in L4 (27) than in L5 (4).
The absolute scaling for the Neptunian Trojans is unknown and we use the value of the Jovian Trojans even though in can be expected that the Neptunian asteroids are much more numerous \citep{Sheppard2006_NeptunianTrojans}.
While this may certainly be due to a bias in the detection efficiency, the ratio leans towards the same direction as the Jovian Trojans.
The estimated parameters for the Neptunian Trojans are listed in Tab.\,\ref{tab:JT_fit_parameters}.

\subsection{Kuiper Belt Objects}\label{sec:KBOs}
Everything between the Neptunian Trojans up to the Oort Cloud, we collect into one accumulation of Trans Neptunian Objects, most of which fall into the category of Kuiper Belt Objects.
Most of the Kuiper Belt Objects are found between 30 and 50\,AU, with a strong peak around 40\,AU.
Their inclinations range up to $50^\circ$ with outliers up to $90^\circ$ (polar orbits) and beyond (retrograde orbits).
Their eccentricities show the full range between $0$ and $<1$ with a large fraction below $0.5$.

We bin the Kuiper Belt Object data into $40 \times 40 \times 22 = 35,200$ $xyz$-bins in the intervals $\left[-90,90\right]$\,AU, $\left[-90,90\right]$\,AU, and $\left[-45,45\right]$\,AU, for a 3D-bin size of $\Delta V = \Delta x \times \Delta y \times \Delta z = 4.5\,\mathrm{AU} \times 4.5\,\mathrm{AU} \times 4.091\,\mathrm{AU} = 82.841\,\mathrm{AU^3}$.
Even though there are only $\sim 4000$ objects, we find that we need three Gaussian tori to fit the Kuiper Belt Object distribution well.
This is due to the wings of the distribution in all directions with extreme outliers.
The fitted parameters for the Kuiper Belt Object density function are listed in Tab.\,\ref{tab:KBO_fit_parameters}.

\begin{table}[h]
	\begin{center}
		\centering
		\caption{Same as Tab.\,\ref{tab:Main Belt Asteroid_fit_parameters} but for Kuiper Belt Asteroids. The total distribution is described by the sum of three, 1--3, tori. Here, the units for $\rho_0$ are $10^{-2}\,\mathrm{AU^{-3}}$.}\label{tab:KBO_fit_parameters}%
		\begin{tabular}{l || r |rrrrr}
			\hline
			
			Group & $N_{\rm obj}$ & $\rho_0$ & $R$ & $\Sigma$ & $r$ & $\sigma$ \\
			
			\hline
			KBO 1 & $333$ & $0.37$ & $10.74$ & $27.88$ & $-1.34$ & $11.62$ \\
			KBO 2 & $1767$ & $3.04$ & $35.73$ & $4.70$ & $-2.47$ & $8.77$ \\
			KBO 3 & $1870$ & $11.50$ & $42.54$ & $3.61$ & $0.33$ & $2.68$ \\
			\hline
		\end{tabular}
	\end{center}
\end{table}

\subsection{Oort Cloud}\label{sec:oort_cloud}
The Oort Cloud is a hypothetical accumulation of asteroids nearly-spherically symmetric around the Sun extending half-way to the next star system \citep{Emelyanenko2007_OortCloud}.
Since the Oort cloud is only theorised but shows some observational evidence, for example from single comets \citep{Hills1981_HillsCloud,Bailey1988_HillsCloud,Emelyanenko2007_OortCloud,Licandro2019_OortCloud}, we will briefly summarise its extent and discuss it in terms of a possible $\gamma$-ray foreground.

According to the \textit{Encyclopedia of the Solar System} \citep{Weissman1999_SolarSystem}, the Oort Cloud might range from somewhere between 2000--5000\,AU up to 50000--100000\,AU (0.24--0.48\,pc).
Only a few comets might have orbital elements that would place their origin in the Oort Cloud, so that we cannot fit a density distribution.
Instead, we assume a spherical Gaussian shell, Eq.\,(\ref{eq:Gaussian_shell}), centred at 40000\,AU with a width of 8000\,AU.
The amplitude of the sphere is unknown and we will use arguments from \citet{Moskalenko2008_GRalbedoSS} to scale the flux in relation to other asteroid accumulations.
While these numbers are somewhat arbitrary, they incorporate the literature values and describe a useful zero-order estimate.
In fact, the total $\gamma$-ray albedo flux will depend mostly on the near edge of the cloud and its thickness, and only a little on its shape.
The time variability of this emission is also expected to be marginal on the sub-percent level (see Sect.\,\ref{sec:time_variability_Earth}).

\begin{figure*}[!ht]%
	\centering
	\includegraphics[width=1.0\textwidth]{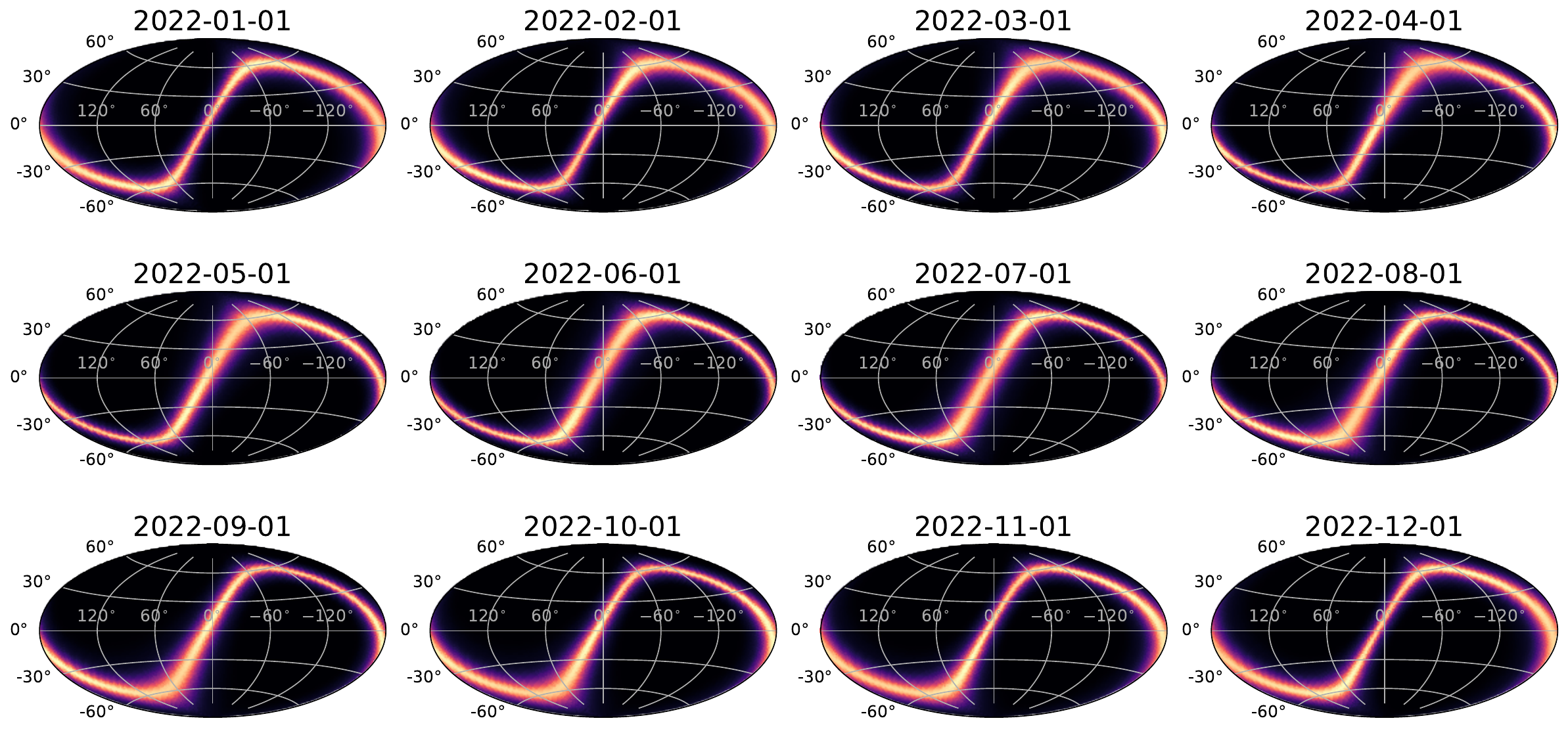}
	\caption{Variation of the Main Belt Asteroid $\gamma$-ray albedo in the course of one year. Shown are the monthly appearance of the Main Belt Asteroids in Galactic coordinates. An animated version of this figure is available here: \href{https://www.physik.uni-wuerzburg.de/astro/mitarbeiter/ag-siegert/Solar-system-gamma-ray-albedo/}{Link}.}
	\label{fig:Main Belt Asteroid_variation}
\end{figure*}

\section{Line-of-sight effects from a moving observer}\label{sec:los}
The density distributions from the previous sections are used to mimic the quasi-diffuse emission that can be expected from the numerous objects along a line of sight.
On average we assume that every asteroid will contribute to the emissivity since the central density of \emph{detected} asteroids in the Main Belt, for example, is about $10^5\,\mathrm{AU^{-3}}$ with a radial size of $\gtrsim 1$\,km.
This gives a ratio of $\sim 10^{-22}$, so that nearly no asteroid is blocking the emission of another asteroid.
In reality, the average density might be much larger, and the average asteroid size much smaller, as the population of asteroids below $\approx 1$\,km is basically unknown.
Assuming a Dohnanyi cascade for the collisional production of smaller pieces from larger ones \citep{Dohnanyi1969_asteroids}, the number of asteroids in the Main Belt could range between $10^{13}$--$10^{14}$ with sizes from 100\,cm to $5 \times 10^{7}$\,cm (size of Ceres).
This would obtain an average number density of $\approx 10^{12}\,\mathrm{AU^{-3}}$, given an effective volume of the Main Belt of $\approx 18\,\mathrm{AU^3}$ when modelled simply with one Gaussian torus, and therefore increase the central density in a similar manner by up to seven orders of magnitude.
The ratio of asteroids per volume is decreased, but the space from one object to another is still large compared to the objects themselves.
We therefore consider no self-blocking of asteroids.
This means that the densities of asteroids from Sect.\,\ref{sec:modelling} can serve as surrogates (first-order proxies) for the expected emissivity,
\begin{equation}
	\rho(x,y,z) \rightarrow \epsilon(x,y,z)\mrm{,}
	\label{eq:emissivity_proxy}
\end{equation}
in units of $\mathrm{ph\,cm^{-3}\,s^{-1}}$ or, equivalently, $\mathrm{erg\,cm^{-3}\,s^{-1}}$.
The $\gamma$-ray albedo flux depends mostly on the surface area of the asteroids (and to some extend on mass and composition) so that a linear scaling of the density will be adapted for the line-of-sight integration.

The variation in time occurs due to the relative motion of asteroid accumulations and the observer.
Since the observer is mainly placed at or around Earth, the variability will always show minimal periods of one sidereal year.
Even if the accumulation is symmetric around the Sun, such as asteroid belts (tori), the motion of Earth let different parts of the belts appear closer and further away throughout one year.
If the accumulation is concentrated to a specific position relative to a planet, such as trojans, there will be apparent epicycles of diffuse emission according to the interplay of the two planets' periods.

\subsection{Line-of-sight integration}\label{sec:los_integral}
In this work, we perform the line-of-sight integration in the (heliocentric) ecliptic frame ($xy$-plane of the Solar System), and transform the frame to Galactic or equatorial for astronomical reference.
The tool we use for the line-of-sight integration is available here\footnote{\url{to_be_added_after_review}}, and include more and general emissivity models, such as dark matter halos, doubly exponential disks, or the \citet{Freudenreich1998_BoxyBulge_COBE} boxy bulge model.
Here, we restrict ourselves to Solar System objects.

In general, the line of sight of an observer is defined by the straight line, $\mathscr{S}$, 
\begin{eqnarray}
	\mathscr{S}: \vec{s}_0(t) + s \cdot \hat{e}_r = \begin{pmatrix}x_0(t) \\ y_0(t) \\ z_0(t)\end{pmatrix} + s \cdot \begin{pmatrix} \cos\phi \cos\theta \\ \sin\phi \cos\theta \\ \sin\theta \end{pmatrix}\mathrm{,}
	\label{eq:los_observer}
\end{eqnarray}
\noindent where $\vec{s}_0(t)$ is the time variable position of the observer, $\hat{e}_r$ is the unit vector in radial direction, and $s \in \mathbb{R}_{0}^+$ the running variable from the point of the observer to infinity.
We define $s$ to be strictly positive (and zero), so that we allow the spherical angles to run in all directions, $\phi \in \left[-180^\circ,+180^\circ\right)$ and $\theta \in \left[-90^\circ, +90^\circ\right]$.
In Eq.\,(\ref{eq:los_observer}), the Cartesian coordinates of $\vec{s}_0(t)$ are taken with respect to the Sun which is at the origin of the coordinate system, $\vec{s}_\odot = (0,0,0)^{\rm T}$.
The $x$-direction is towards the vernal equinox and therefore stays the same -- only the observer position is changing with time.

\begin{figure*}[!t]%
	\centering
	\includegraphics[width=1.0\textwidth]{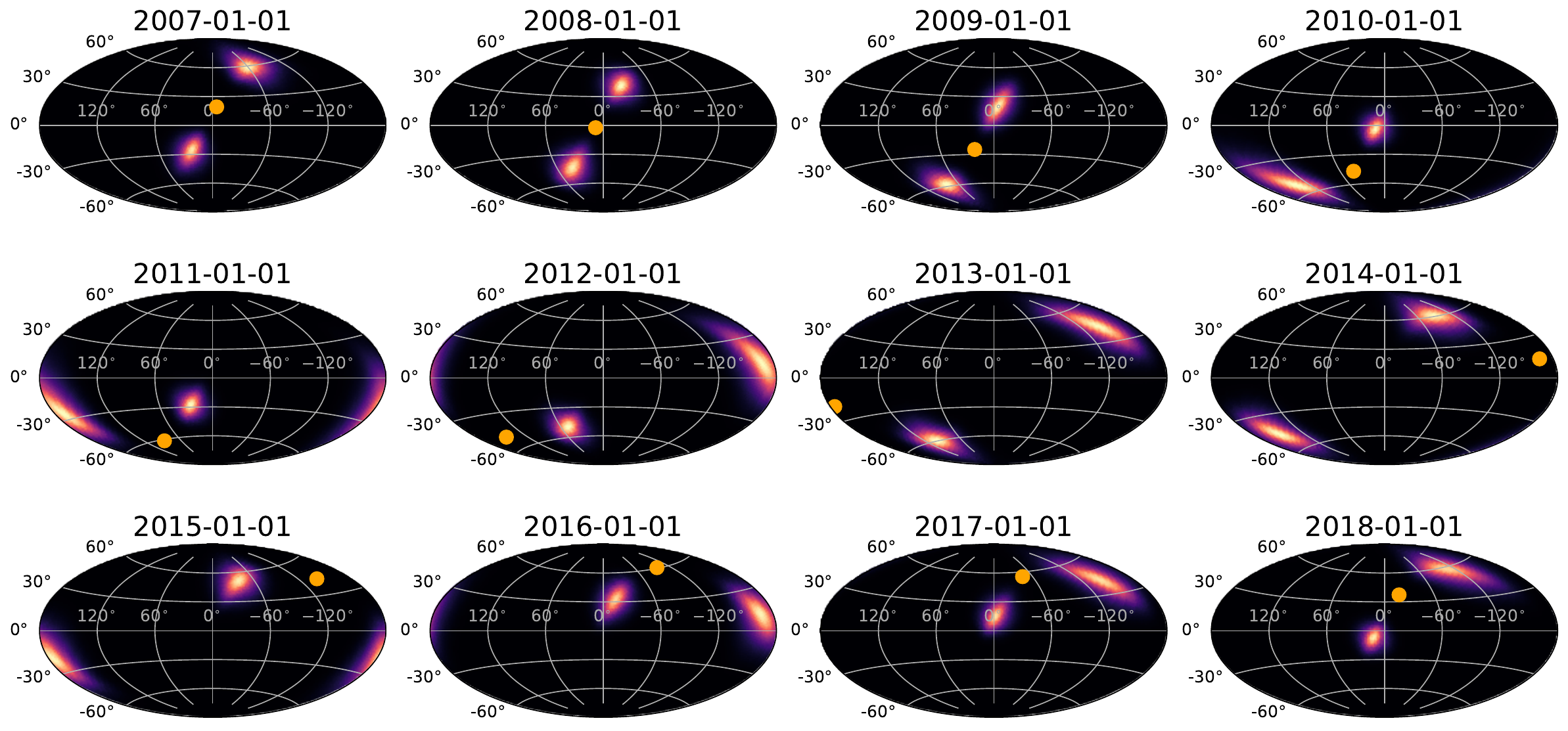}
	\caption{Variation of the Jovian Trojan $\gamma$-ray albedo in the course of twelve years (about one Jupiter orbit). Shown are the yearly appearance of the Jovian Trojans in Galactic coordinates, together with the position of Jupiter indicated as orange dot. An animated version of this figure together with Fig.\,\ref{fig:NTS_variation} is available here: \href{https://www.physik.uni-wuerzburg.de/astro/mitarbeiter/ag-siegert/Solar-system-gamma-ray-albedo/}{Link}.}
	\label{fig:JTS_variation}
\end{figure*}

The line-of-sight integration,
%
\begin{eqnarray}
	F(\phi,\theta,t) & = &  \frac{1}{4\pi\,\mathrm{sr}}\int_{0}^{\infty}\,ds\,\epsilon(s) = \nonumber\\ 
	& = & \frac{1}{4\pi\,\mathrm{sr}}\int_{0}^{\infty}\,ds\,\epsilon(x_0(t) + s \cdot \cos\phi \cos\theta, \nonumber\\ 
	& & y_0(t) + s \cdot \sin\phi \cos\theta, z_0(t) + s \cdot \sin\theta)\mrm{,}
	\label{eq:LOS_integration}
\end{eqnarray}
%
determines the flux per unit solid angle in units of $\mathrm{ph\,cm^{-2}\,s^{-1}\,sr^{-1}}$ or $\mathrm{erg\,cm^{-2}\,s^{-1}\,sr^{-1}}$ as a function of time of the observation.
The corresponding luminosity, in units of $\mathrm{ph\,s^{-1}}$ or $\mathrm{erg\,s^{-1}}$, is calculated by
%
\begin{eqnarray}
	L & = & \int_{4\pi}\,d\Omega\int_{0}^{\infty}\,ds\,s^2\epsilon(s) = \nonumber\\ 
	& = &  \int_{-\pi}^{+\pi}\,d\phi\int_{-\pi/2}^{+\pi/2}\,d\theta\sin\theta\int_{0}^{\infty}\,ds\,s^2\epsilon(x_0(t) +  \nonumber\\ 
	& + & s \cdot \cos\phi \cos\theta, y_0(t) + s \cdot \sin\phi \cos\theta, z_0(t) + s \cdot \sin\theta)\mrm{,}
	\label{eq:LOS_luminosity}
\end{eqnarray}
%
which is equivalent to a volume integral, weighted with the emissivity.
In the case of the asteroid densities per unit volume, this `luminosity' equals the number of \emph{detected} asteroids, and 
\begin{equation}
	N(\phi,\theta) = \int_{0}^{\infty}\,ds\,s^2 \rho(s,\phi,\theta)
	\label{eq:number_per_area}
\end{equation}
is the number of \emph{detected} asteroids per unit solid angle.
Assuming only the known asteroids, Eq\,(\ref{eq:number_per_area}) provides a lower limit on how diffuse an accumulation will appear, that is, at what angular resolution the emission might be point-like (one or less asteroid per pixel).
As an example, we calculate the `luminosity map' of Main Belt Asteroids in a $1^\circ \times 1^\circ$ pixellised map.
This gives a contribution of at least 70 known asteroids (per square-degree) in the ecliptic plane.
These would be point like sources with an average angular (projected) distance of much smaller than $1^\circ$, so that the emission will appear diffuse down to a scale of roughly $0.1^\circ$.
We emphasise again that this applies for the \emph{detected} asteroids in the Main Belt Asteroid, and may be even true on a much smaller angular scale due to the unknown but estimatable population of sub-km asteroids.
We therefore assume in the following that all the emission will be diffuse and smooth on at least an angular scale of $0.1^\circ$.

In practise, Eqs.\,(\ref{eq:LOS_integration})--(\ref{eq:number_per_area}) have no analytical solutions, except for a few cases of solid, homogeneously filled spheres or tori, for example.
We therefore approximate the integrals by Riemann sums on a grid of solid angles (pixels) times a grid of the running variable $s$, so that
\begin{equation}
	F(\phi_j,\theta_k,t) \approx \frac{1}{4\pi\,\mrm{sr}} \sum_{i = i_{\rm min}}^{i_{\rm max}} \epsilon(s_i,\phi_j,\theta_k,t) \Delta s\mrm{,}
	\label{eq:LOS_approx}
\end{equation}
where $i_{\rm min}$ and $i_{\rm max} = i_{\rm min} + n \cdot i$, respectively with $n$ the number of line elements along $s$, are chosen appropriately for the emissivity profile, that is, so that a reasonable accuracy is achieved.
A rejection sampling algorithm may be used to check the accuracy but which might take a much longer computation time.
%
Likewise, we approximate the angle integral in Eq.\,(\ref{eq:LOS_luminosity}) by the a Riemann sum with a solid angle element
\begin{equation}
	d\Omega \approx \Delta\Omega_{jk} = \left[\sin\left(\theta_k + \Delta\omega_{jk}/2\right) - \sin\left(\theta_k - \Delta\omega_{jk}/2\right)\right]\Delta\omega_{jk}\mrm{,}
	\label{eq:solid_angle_approx}
\end{equation}
where $\Delta\Omega_{jk}$ is the solid angle centred around $\phi_j$ and $\theta_k$ with a pixel size of $\Delta\omega_{jk} \equiv \Delta\phi_j$.
We note that in this spherical rectangular grid, the longitudinal pixel size is constant whereas the latitudinal pixel size shrinks according to Eq.\,(\ref{eq:solid_angle_approx}).
The sum over all $\Delta\Omega_{jk}$ should be and is approximately $4\pi\,\mrm{sr}$.
%

\begin{figure*}[!t]%
	\centering
	\includegraphics[width=1.0\textwidth]{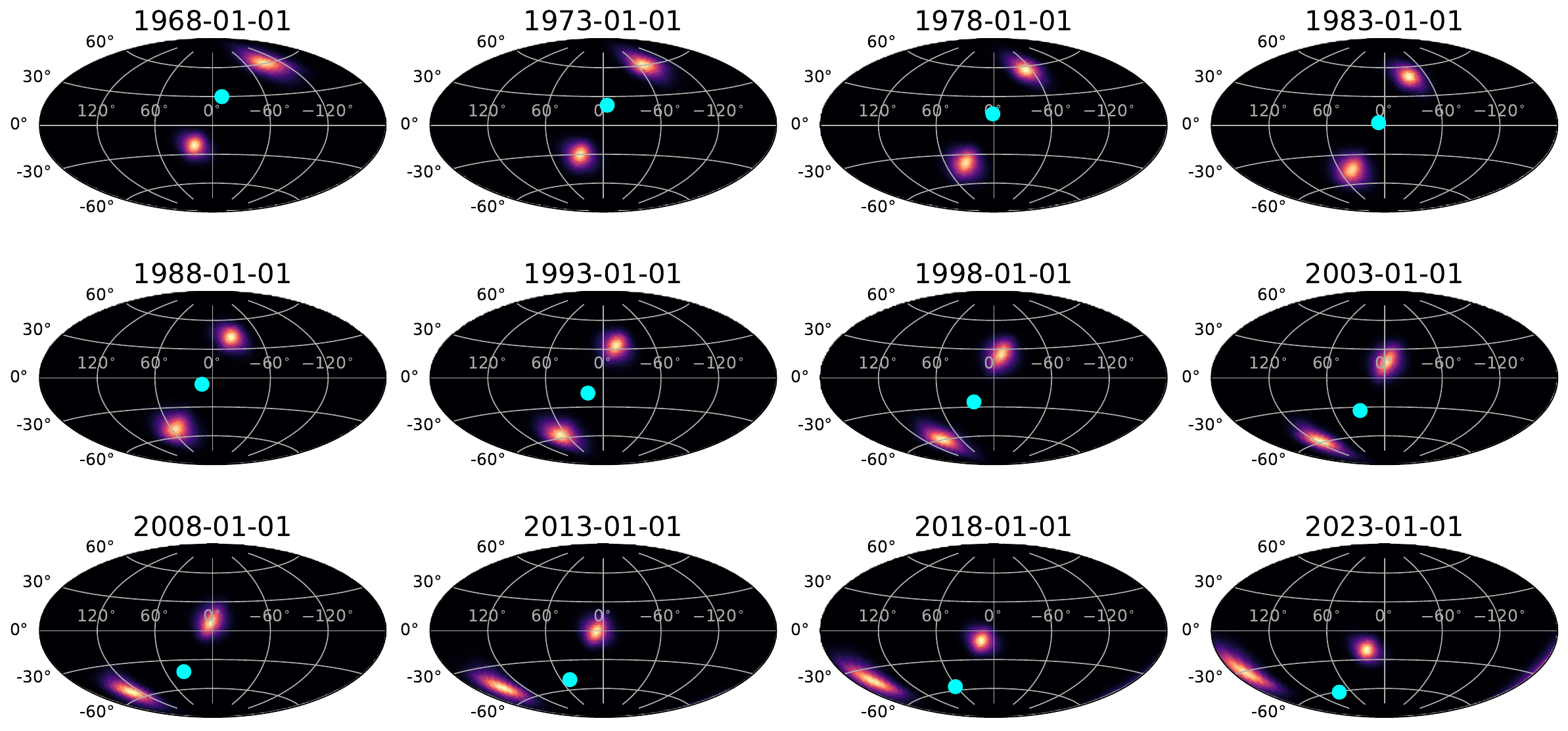}
	\caption{Variation of the Neptunian Trojan $\gamma$-ray albedo in the course of 55 years (about one third of a Neptune orbit). Shown are the appearances every five years of the Neptunian Trojans in Galactic coordinates, together with the position of Neptune indicated as cyan dot. An animated version of this figure together with Fig.\,\ref{fig:JTS_variation} is available here: \href{https://www.physik.uni-wuerzburg.de/astro/mitarbeiter/ag-siegert/Solar-system-gamma-ray-albedo/}{Link}.}
	\label{fig:NTS_variation}
\end{figure*}

\subsection{Brightness variations within an Earth-year}\label{sec:time_variability_Earth}
Variations on the timescale of an Earth-year (sidereal year) originate from the relative motion of the observer on Earth with respect to other objects in the Solar System.
This holds true even if the density structure is symmetric around the Sun.
The most extreme variation within one year stems from the Main Belt Asteroids as the relative distance of Earth to specific points inside the belt is 1.8--3.8\,AU (mean orbit around 2.8\,AU).
%
%
Based on the above minimum and maximum distance, the largest flux ratio along the Main Belt Asteroids could be expected to be around $(1.8/3.8)^2 \approx 0.22$.
However, the line-of-sight integration with the given density profile, Sect.\,\ref{sec:Main Belt Asteroid_complete}, obtains a largest flux ratio of $\approx 0.45$ because the 3D-emissivity has a ($3\sigma$) thickness of $\approx 1$\,AU equally at all distances, which impacts directly the diffuse flux.
Fig.\,\ref{fig:Main Belt Asteroid_variation} shows the variation of the Main Belt Asteroids for each month in the Earth year 2022.
Due to the nearly-circular orbit of Earth, the enhancement of the Main Belt Asteroid flux along the ecliptic appear at regular time intervals with the same relative increase.
The cumulative effect of the apparent variation of the Main Belt Asteroid albedo is a broadening of the total ecliptic band.
While the 1$\sigma$-width of the emission varies between $5.5^\circ$ and $12.5^\circ$ at an individual point in time, the cumulative width is $10^\circ$.
Thus when the ecliptic is taken into account in $\gamma$-ray data analyses, each (pointed) observation should obtain its own emission template as a temporally averaged template will be too broad or too narrow, so that such a false input map may mimic point like emission in certain regions due to the residual, not accounted for, emission.

The same effect appears for the Kuiper Belt Objects but since the distances are much larger (30--50\,AU), the flux variation is small around 1\,\%.
The general difference between the Main Belt Asteroid and Kuiper Belt Objects albedo is the vertical extent.
While the Main Belt Asteroids are concentrated roughly within a torus of $0.8$\,AU width, the Kuiper Belt Objects are confined within $12$\,AU vertically.
Given the average distances, this means that the Kuiper Belt Objects populate slightly higher ecliptic latitudes (see also Sect.\,\ref{sec:full_albedo_model}).

The Jovian and Neptunian Trojans move about $30^\circ$ and $2^\circ$, respectively, during one year as seen from Earth.
While this may mimic some more extended sources, their relative motion to Earth is more important on the decadal time scale which is also typical for diffuse emission measurements in the $\gamma$-ray range.
Due to their orbital periods of $11.86$\,yr and $164.79$\,yr, respectively, the trojans appear concentrated and fixed at a celestial location for shorter amounts of time, such as for $\gamma$-ray transient observations (typically less than a few days) or even for the canonical observation time of $1$\,Ms.
Within this time scale, the motion of the Jovian Trojans is $<1^\circ$ and that of the Neptunian Trojans is $<0.1^\circ$, so that for soft $\gamma$-ray telescopes with a typical angular resolution of a few degrees, these variations are irrelevant.
For high-energy $\gamma$-ray telescopes, this variation should be taken into account as it might mimic extended emission due to the combination of large data sets.
Likewise, the Oort cloud would also show a small anisotropy, probably on the level of $10^{-10}$ or less, and we therefore ignore its variation in the following.

\begin{figure*}[!t]%
	\centering
	\includegraphics[width=1.0\textwidth]{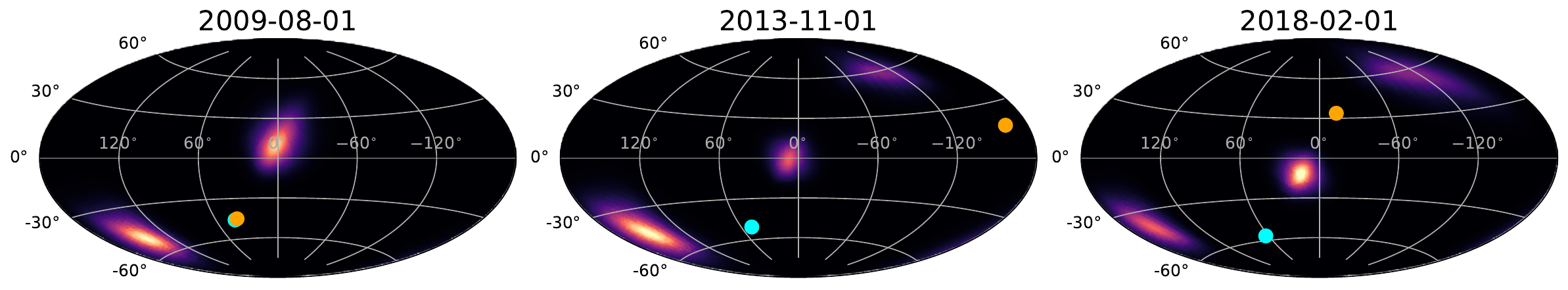}
	\caption{Overlaps of Jovian and Neptunian Trojans in full conjunction (left) and $\pm 120^\circ$ trailing (middle and right). Shown are the appearances of the Jovian Trojans and Neptunian Trojans every $4.25$ years in Galactic coordinates, together with the position of Neptune and Jupiter indicated as cyan and orange dots, respectively. The total fluxes of the Jovian Trojans and Neptunian Trojans are assumed to be identical. An animated version of this figure as a combination of Figs.\,\ref{fig:JTS_variation} and \ref{fig:NTS_variation} is available here: \href{https://www.physik.uni-wuerzburg.de/astro/mitarbeiter/ag-siegert/Solar-system-gamma-ray-albedo/}{Link}.}
	\label{fig:JT_NT_overlap}
\end{figure*}

\subsection{Brightness variations on the decade time scale}\label{sec:time_variability_decade}
For very long observations -- the data sets of $\gamma$-ray telescopes can easily reach more than ten years -- the relative motions of all asteroid accumulations with respect to Earth become important.
When Jupiter appears to move in epicycles due to the approach and recession of Earth with respect to the planet, its trojans show the same behaviour because the accumulation as a whole is on a similar orbit.
In fact, the approaching trojans appear brighter due to the line-of-sight integration, whereas the receding ones appear dimmer and more extended.
In Fig.\,\ref{fig:JTS_variation}, we show the variation of the Jovian Trojans on a timescale of 12\,yr (roughly one Jupiter orbit).

The Neptunian Trojans move by only $2^\circ\,\mrm{yr^{-1}}$, so that their impact becomes apparent for much longer timescales.
Since the number and extent of the Neptunian Trojans are only extrapolated from the Jovian Trojans, their appearance is in general very similar.
Human $\gamma$-ray observations started in the 1960s with first detections of hundreds of photons with Explorer 11 (1961) and OSO3 (1967).
Since then, the Neptunian Trojans moved about $120^\circ$ across the sky in the ecliptic, not even completing one orbit of Neptune.
Their cumulative effect mimics, intriguingly, emission near the Galactic Centre and bulge region and should therefore be taken into account when analysing these regions in particular (see also Sect.\,\ref{sec:GCE_mimic}).
We show the expected appearance of the $\gamma$-ray albedo from the Neptunian Trojans in Fig.\,\ref{fig:NTS_variation} for a timescale of the last 55\,yr, that is, when $\gamma$-ray observations became feasible and meaningful.

The cumulative effect of both trojan accumulations is an emission `strip' along the ecliptic with varying intensities.
Depending on the actual observation time, this may either include the entire ecliptic or only parts of it.
Since the trojans also perform epicyclic motions, longer integration times do not automatically smear out the signal towards a Main-Belt-Asteroid-like structure, but may in fact enhance certain regions in the sky at particular times.
Especially if the trojans from Jupiter and Neptune overlap, either completely aligned when the planets are in conjunction (every $12.78$\,yr), or the trailing trojans of Neptune are caught up by the leading trojans of Jupiter, or vice versa (also both every $12.78$\,yr).
This means, roughly every $4.26$\,yr, the Jovian and Neptunian Trojans show some partial overlap and the flux from the overlapping direction is increased accordingly.
We illustrate this behaviour in Fig.\,\ref{fig:JT_NT_overlap}.
The effect of enhanced emission can be further pushed if the yearly Main Belt Asteroid enhancement is considered (see Sect.\,\ref{sec:full_albedo_model}).

\begin{figure*}[!t]%
	\centering
	\includegraphics[width=1.0\textwidth]{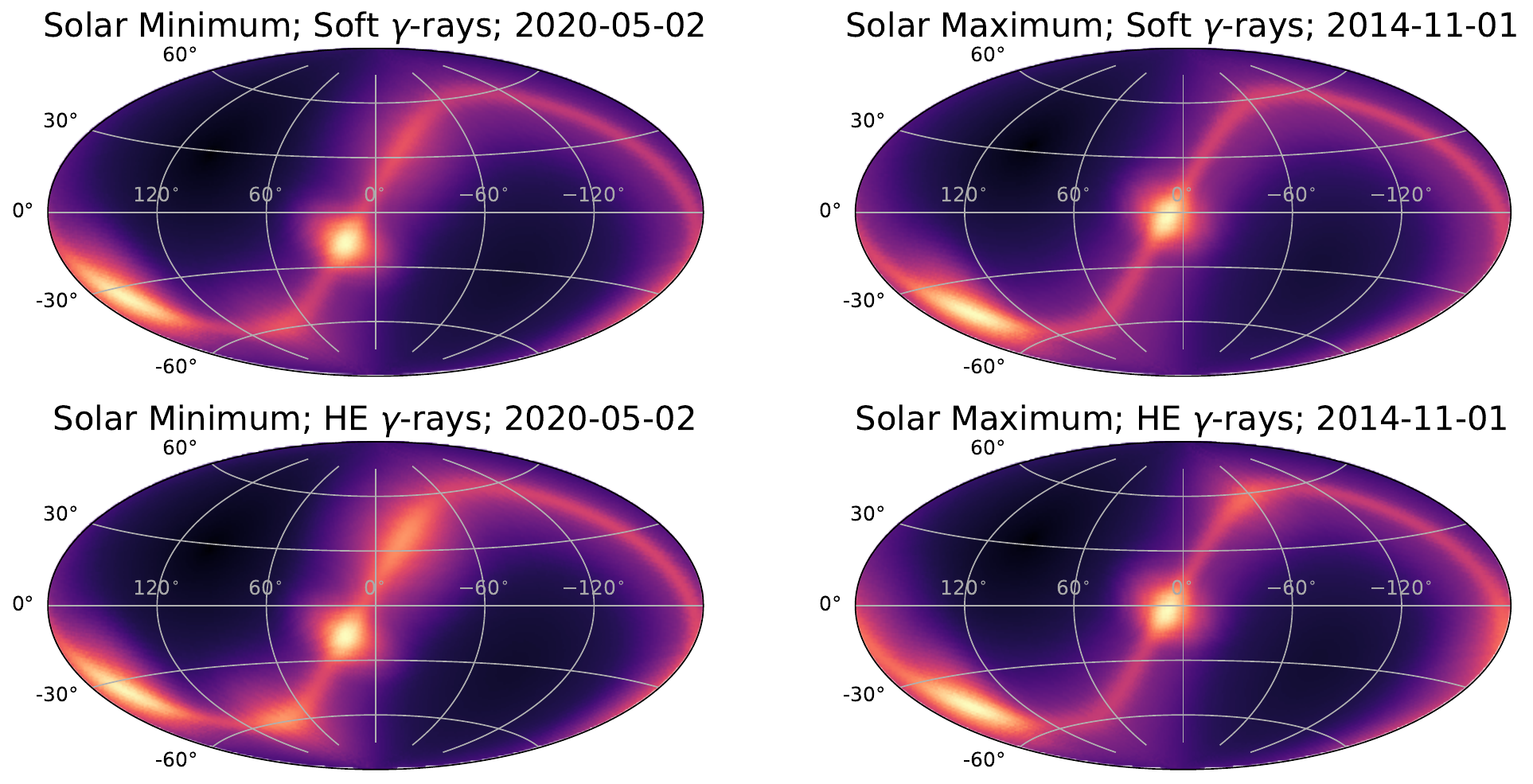}
	\caption{Full model of the diffuse time-variable Solar System albedo emission. Shown are the expected all-sky maps during a Solar minimum (left) and a Solar maximum (right), for soft $\gamma$-ray photons ($E_{\gamma} \lesssim 10$\,MeV) and high-energy $\gamma$-ray photons ($E_{\gamma} \gtrsim 10$\,MeV) in Galactic coordinates. During a Solar minimum, all components have a similar all-sky flux for HE photons, and the attenuation for the inner components (mainly Main Belt Asteroids and Jovian Trojans) is also weak for soft $\gamma$-rays. During a Solar maximum, mainly the Kuiper Belt Objects and Neptunian Trojans are visible and outshine the inner components. The colour is scaled by the cube-root to emphasise on the weaker components. An animated version of this figure without the Solar Cycle is available here: \href{https://www.physik.uni-wuerzburg.de/astro/mitarbeiter/ag-siegert/Solar-system-gamma-ray-albedo/}{Link}.}
	\label{fig:total_model_maps}
\end{figure*}

\subsection{Full small Solar System body albedo emission}\label{sec:full_albedo_model}
For a full model of how the time variable Solar System albedo including Main Belt Asteroids, Jovian Trojans, Neptunian Trojans, Kuiper Belt Objects, and the Oort Cloud, behaves, the total number of asteroids for each accumulation is required.
As described in Sect.\,\ref{sec:los_integral}, the number of asteroids can serve as a proxy for the expected luminosity.
While the total mass of these accumulations is dominated by the largest objects (e.g., Ceres in the Main Belt carrying $\approx 30$\,\% of the total mass \citep{Krasinsky2002_Ceres}, or Pluto in the Kuiper Belt carrying $\approx 18$\,\% of the total mass), the luminosity is dominated by the smallest objects, whose numbers are difficult to estimate.
Furthermore, the expected $\gamma$-ray flux will depend on the cosmic-ray flux as a function of heliocentric distance (rising with distance) as well as the Solar modulation potential (falling with distance).
This means that the same body will be exposed to a higher dose of cosmic rays at larger distances and therefore its $\gamma$-ray albedo will be stronger, that is, its luminosity will be higher.

We will consider the estimates by \citet{Moskalenko2008_GRalbedoSS} for the total number of asteroids and their luminosity to study the possible appearance of the total model.
%
%
%
%
%
In their work, they estimate the total number of Main Belt Asteroids greater than $1$\,m to be $N_{\rm MBA} \approx 5 \times 10^{11}$ with a power law index of $n=2$ for the number-size-distribution, with about one order of magnitude uncertainty.
We will not extrapolate to objects to smaller radii, even though those might be much more numerous, because the albedo spectrum might be (much) weaker compared to larger objects as the cosmic-ray cascade for smaller objects will not entirely develop \citep{Moskalenko2007_GRalbedoMoon,Moskalenko2009_CGB_gammaray_albedo}.
\citet{Moskalenko2008_GRalbedoSS} determine that the total number of Jovian Trojans is similar to that of Main Belt Asteroids as their size distributions are similar, that is $N_{\rm JT} \approx 5 \times 10^{11}$.
The number of Neptunian Trojans might outnumber the Jovian Trojans by a factor of 10 for large ($\gtrsim 80$\,km) objects, and their size distribution may follow $n=2.5$, so that the number of Neptunian Trojans may be around $N_{\rm NT} \approx 10^{13}$.
The number of Kuiper Belt Objectss greater than $1$\,m is quoted as $N_{\rm KBO} \approx 5 \times 10^{16}$, with $n=2.5$, and 2.5 orders of magnitude uncertainty.
For the Oort Cloud, \citet{Moskalenko2009_CGB_gammaray_albedo} estimate the number of $>1$\,km sized objects to be $(3$--$14) \times 10^{12}$.
If we apply an $n=2.5$ power law to estimate the total number down to $1$\,m \citep{Dohnanyi1969_asteroids}, we obtain $N_{\rm OC} \approx 3 \times 10^{20}$.

The relative weights from the Solar modulated cosmic-ray flux, $F_{\rm CR}(E,r,\Phi)$ is calculated from Eqs.\,(7) and (8) of \citet{Moskalenko2009_CGB_gammaray_albedo} and Eq.\,(13) of \citet{Moskalenko2008_GRalbedoSS}, with the Local Interstellar Cosmic-ray spectrum by \citet{Vos2015_LICRS}.
Values beyond $\approx 122$\,AU are considered to be exactly this spectrum.
We use the cosmic-ray flux at a distance to the Sun as a linear multiplicator for the luminosity, $L \propto N \times F_{\rm CR}$, and consider particles in the range $10^2$--$10^4\,\mathrm{MeV\,nucleon^{-1}}$.
For $10$\,GeV particles, the Solar potential has almost no impact and the relative increase from 2.8\,AU (Main Belt Asteroids) to 40000\,AU (Oort Cloud) is at most 60\,\% for high modulation ($\Phi = 1500$\,MV; Solar maximum at 1\,AU) and merely 20\,\% for moderate modulation ($\Phi = 500$\,MV).
For $100$\,MeV particles, however, the relative increase at the Oort Cloud is up to $10^5$ for high and up to $10^3$ for moderate modulation.
In Tab.\,\ref{tab:cosmic_ray_fluxes}, we show the normalisations, up to a constant multiplicative factor that includes the actual spectral shape, for different Solar modulation potentials at the (effective) distances of the asteroid accumulations for the two cases of 100\,MeV and 10\,GeV protons.
While the low-energy particles are mostly responsible for the soft $\gamma$-ray part of the expected albedo spectrum ($E_{\gamma} \lesssim 10$\,MeV), the higher energy particles are creating the `hard' $\gamma$-ray part ($E_{\gamma} \gtrsim 10$\,MeV) including the pion peak.
In fact, the Solar modulation potential as a function of time will invoke another source of variability on human time scales.
We use a sinusoidal dependence with a period of $P = 11$\,yr to mimic the Solar cycle,
\begin{equation}
	\Phi(t) = \Phi_0 \left[\sin\left(\frac{2\pi}{P}t + \omega_0\right) + 1.2\right]\mrm{,}
	\label{eq:modulation_potential_time}
\end{equation}
with $t$ in Julian years, $\Phi_0 = 600$\,MV and $\omega_0 = 0.52$.
We note that the Solar modulation potential has a much more complex shape, that the peak potential varies between cycles, and that the cycles may in fact have a duration between 8 and 14 years \citep[e.g.,][]{Barnard2011_Solarweather}.
For the purpose of this work, this rough estimate is deemed sufficient as the intrinsic luminosity, as given by the number of asteroids in an accumulation, is uncertain by at least an order of magnitude (see Tab.\,\ref{tab:full_sssb_models}).
For the soft $\gamma$-ray flux, this results in additional variations of 3--5 orders of magnitude throughout the Solar cycle, depending on heliocentric distance.
Higher energies are affected only by a factor of 2 at most.

We normalise the component to each other relatively with the number of asteroids we would obtain from Eq.\,(\ref{eq:LOS_luminosity}), and the relative increase of the cosmic-ray flux with distance from the Sun, modulated with time.
We provide Tab.\,\ref{tab:full_sssb_models} with all normalisations, effective volumes of the chosen density functions, relative luminosities, and (total) flux ratios.
Given the luminosity normalisations, we can calculate the expected emission maps for the entire model for all times.
The models are shown in Fig.\,\ref{fig:total_model_maps}.

\begin{table}[h]
	\begin{center}
		\caption{Relative logarithmic decrease of the Local Interstellar Cosmic-ray flux, $F_{\rm CR}(E,r,\Phi)$. The values are taken with respect to a distance of $r \approx 100$\,AU where the impact of the Solar modulation vanishes, $\lg\left[F_{\rm CR}(E,r,\Phi)/F_{\rm CR}(E,r=100\,\mathrm{AU},\Phi = 0\,\mathrm{MV})\right]$, that is, when the Solar modulation potential tends to zero.}\label{tab:cosmic_ray_fluxes}%
		\begin{tabular}{c||c| r r r r r}
			\hline
			&  & \multicolumn{5}{c}{Distance [AU]} \\
			Energy & $\Phi$ [MV] & 2.4 & 5.5 & 29.4 & 34.4 & $10^4$ \\
			\hline
			\multirow{4}{*}{100\,MeV} & 0 & 0 & 0 & 0 & 0 & 0 \\
			& 500 & -3.2 & -2.7 & -1.5 & -1.4 & 0\\
			& 1000 & -4.3 & -3.8 & -2.4 & -2.2 & 0\\
			& 1500 & -5.0 & -4.5 & -3.0 & -2.7 & 0\\
			\hline
			\multirow{4}{*}{10\,GeV} & 0 & 0 & 0 & 0 & 0 & 0\\
			& 500 & -0.08 & -0.06 & -0.02 & -0.02 & 0\\
			& 1000 & -0.15 & -0.11 & -0.04 & -0.04 & 0\\
			& 1500 & -0.22 & -0.17 & -0.07 & -0.06 & 0\\
			\hline			
		\end{tabular}
	\end{center}
\end{table}

\begin{table*}[t]
	\begin{center}
		\centering
		\caption{Parameters of the total Solar System $\gamma$-ray albedo model for the case without Solar modulation, $\Phi = 0$. From left to right, the parameters are the decadal logarithm of the number of objects larger than $1$\,m, $\lg N_{\rm obj}$, the effective distance of the asteroid accumulation, $\tilde{d}$, in units of AU, the effective volume of the accumulation, $V_{\rm eff}$, in units of $\mathrm{AU^3}$, the decadal logarithm of the `flux' according to Eq.\,(\ref{eq:LOS_integration}), $\lg \tilde{F}_{\rm obj}$, in arbitrary relative units, the all-sky flux of an accumulation with respect to the Main Belt Asteroid all-sky flux, $\tilde{F}_{\rm obj} / \tilde{F}_{\rm Main Belt Asteroid}$, and the conversion factor from luminosity to flux according to Eq.\,(\ref{eq:LOS_luminosity}). The luminosities, and therefore fluxes, are to be scaled by the cosmic-ray flux as a function of heliocentric distance and Solar modulation potential according to Tab.\,\ref{tab:cosmic_ray_fluxes}.}\label{tab:full_sssb_models}%
		\begin{tabular}{l || r |rrrrr}
			\hline
			\\
			Group & $\lg N_{\rm obj}$  & $\tilde{d}$ & $V_{\rm eff}$ & $\lg \tilde{F}_{\rm obj}$ & $\tilde{F}_{\rm obj} / \tilde{F}_{\rm Main Belt Asteroid}$ & $\tilde{L}_{\rm obj} / \tilde{F}_{\rm obj}$ \\
			
			\hline
			Main Belt Asteroid & $11.7 \pm 1.0$ & $2.4$ & $73$ & $-1.9 \pm 1.0$ & $1.0$\,(fix) & $7.2 \cdot 10^{1}$ \\
			JT & $11.7 \pm 1.0$ & $5.5$ & $16$ & $-2.6 \pm 1.0$ & $0.18^{+2.45}_{-0.16}$ & $3.8 \cdot 10^{2}$ \\
			NT & $13.0 \pm 1.0$ & $29.4$ & $1920$ & $-2.7 \pm 1.0$ & $0.13^{+1.73}_{-0.11}$ & $1.1 \cdot 10^{4}$ \\
			KBO & $16.7 \pm 2.0$ & $34.4$ & $1.6 \cdot 10^5$ & $0.8 \pm 2.0$ & $483^{+48040}_{-434}$ & $1.5 \cdot 10^{4}$ \\
			OC & $20.5 \pm 2.5$ & $4 \cdot 10^4$ & $4.2 \cdot 10^{14}$ & $-1.5 \pm 2.5$ & $2.06^{+648}_{-1.85}$ & $2.1 \cdot 10^{10}$ \\
			\hline
		\end{tabular}
	\end{center}
\end{table*}

As a result this means that, depending on the Solar modulation potential, and given the uncertainties on the absolute number of asteroids in each group, either component may be the brightest at a specific time.
Taking the relative flux values of Tab.\,\ref{tab:full_sssb_models} at face value, the Kuiper Belt Objects are the brightest source of a $\gamma$-ray albedo at Solar minima for both, low- and high-energy photons.
The same is true for Solar maxima, in which case the Kuiper Belt Objects may outshine the remaining sources by 2--3 orders of magnitude.
However, given that especially the total number of Kuiper Belt Objects is uncertain by two orders of magnitude, one can find configurations in which the Kuiper Belt Objects are subdominant and the trojans appear as bright extended sources.
In particular, taking the groups up to the Neptunian Trojans at the upper bound and the Kuiper Belt Objects at the lower bound, the Neptunian Trojans appear as the strongest source at Solar maximum for both, low- and high-energy photons.
At Solar minima, such a configuration lets the Neptunian Trojans appear as the strongest source as well with the Main Belt Asteroids shining dim along the ecliptic for soft $\gamma$-rays.
For `hard' $\gamma$-rays, all sources appear similar in brightness in this case.
We note that a similar all-sky flux does not imply a similar brightness as the sources are distributed across the sky and have different extensions:
The Oort cloud albedo, for example, would show a high flux, but distributed over the whole sky will make the source (and its structure) have only little contributions per unit pixel.
If the number of Neptunian Trojans is much lower, the Jovian Trojans would appear as enhancements of the bright ecliptic due to the Main Belt Asteroids.
For simplicity in the calculations of Fig.\,\ref{fig:total_model_maps}, we assume that all components have the same flux intrinsically when the Solar modulation is ignored.
This assumption is within the uncertainties from Tab.\,\ref{tab:full_sssb_models}.

The uncertainties in the absolute normalisations show that each component may in fact be the strongest, so that for future $\gamma$-ray observations in the sub-GeV bands, each component should be carefully included.
We show one particular case of a possible image artefact which may in fact be due to the cumulative effect of the Jovian Trojans over a long observation period in Sect.\,\ref{sec:OSSE_fountain}.
In addition we show that especially Galactic Centre observations should include this time variable $\gamma$-ray foreground as the Neptunian Trojans can mimic some contribution towards this direction (Sect.\,\ref{sec:GCE_mimic}).

\begin{figure*}[!t]%
	\centering
	\includegraphics[width=1.0\textwidth]{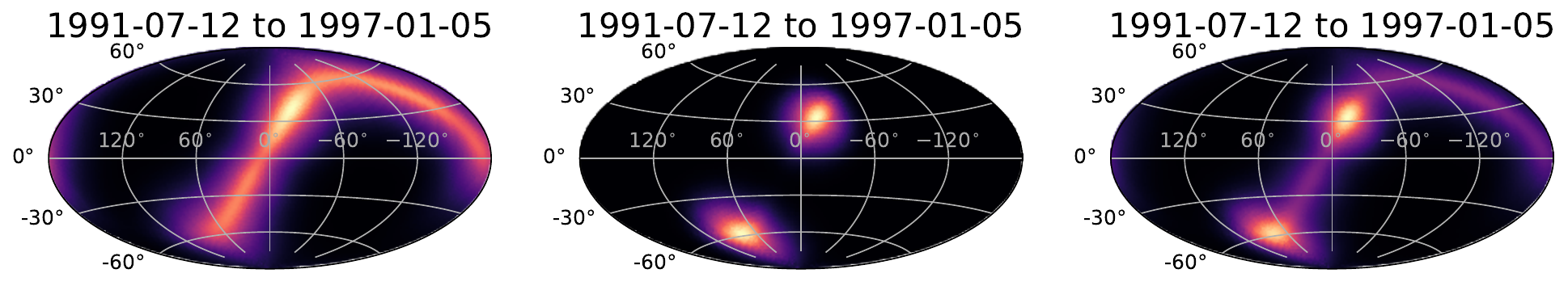}
	\caption{Cumulative emission of Jovian Trojans (left), Neptunian Trojans (middle), and combined emission (right) from 1991-07-12 to 1997-01-05, i.e. observation times of OSSE according to \citet{Purcell1997_511}. Especially the Jovian Trojans appear asymmetric along the ecliptic as the observation time is only half the orbital period of Jupiter of $11.86$\,yr. Furthermore, the apparent epicyclic motion of the trojans as well as the overlap of the emission of the L4 and L5 trojans (every $\approx 4.26$\,yr) create a positive latitude enhancement. The Neptunian Trojans might then add emission above latitudes of $\gtrsim 15^\circ$. The images are scaled by the square-root to emphasise on weak emission.
	}
	\label{fig:JT_NT_OSSE}
\end{figure*}

\begin{figure*}[!t]%
	\centering
	\includegraphics[width=1.0\textwidth]{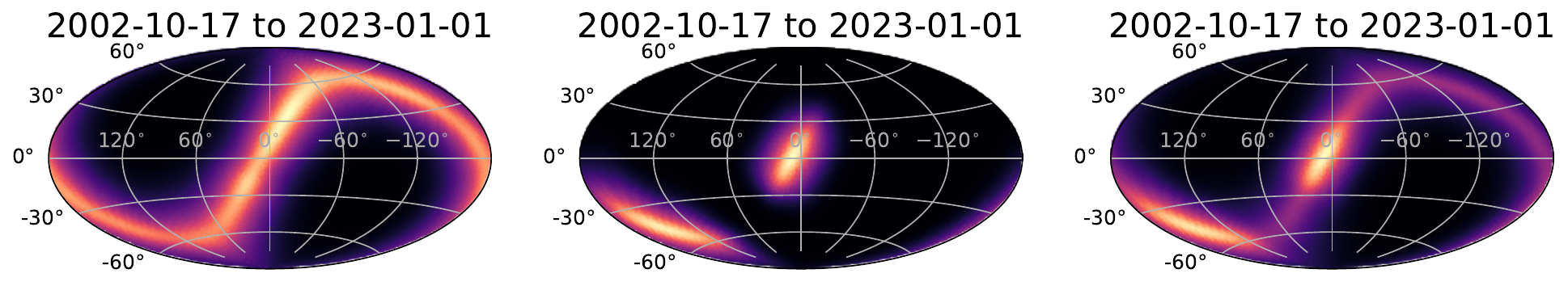}
	\caption{Same as Fig.\,\ref{fig:JT_NT_OSSE} but for the time between 2002-10-17 (launch of the INTEGRAL satellite) and 2023-01-01. Here, the Jovian Trojans revolved about twice around the Sun, but because of the epicyclic motion, the cumulative emission is centred towards the Galactic bulge. The Neptunian Trojans fully overlap with the Galactic bulge and might create a positive longitude enhancement at $0 \lesssim \ell \lesssim +15^\circ$.
	}
	\label{fig:JT_NT_GCE}
\end{figure*}

\section{Apparent image artefacts and enhancements}\label{sec:image_artefacts}
Imaging of $\gamma$-ray data is relying on assumptions of how the $\gamma$-ray sky looks like at any given time.
While image reconstruction algorithms, such as Richardson-Lucy \citep{Richardson1972_RichardsonLucy,Lucy1974_RichardsonLucy}, Maximum Entropy \citep{Narayan1986_MaxEnt}, Singular Value Decomposition \citep[e.g.,][]{Purcell1997_511}, among others, can provide an unbiased representation of the data, the reconstructions will always suffer from the missing data problem which can result in imaging artefacts, serendipitous detections, or apparent transients.
The more proper way of creating $\gamma$-ray images, forward folding of (parametrised) image templates (as a function of energy), may be more biased to the extent that a certain prior belief is included, but their resulting parameters and uncertainties can be straight-forwardly interpreted in terms of fit adequacy and physical meaning.
Both methodologies can lead into misinterpretations of the actual data if the time domain is not considered.
Individual, strong point sources are easily taken into account in such analyses by either studying them separately or simply masking out those times for the longer exposure datasets.
However, if the diffuse emission also appears to change with time as expected from the Solar System albedo, either weak sources may appear or entire large scale structures.
In the following, we will briefly discuss two occasions for which the asteroid accumulations may mimic emission which may either be interpreted as due to image artefacts or astrophysical sources.

\subsection{The OSSE 511\,keV fountain}\label{sec:OSSE_fountain}
Especially MeV reconstructions appear shaky, sometimes with fountain-like features around the Galactic Centre, which is mainly due to the limited number of celestial photons measured compared to the number of pixels defined in the analysis.
One of the possible cases in which the trojans may have mimicked such a feature is the `Galactic 511\,keV fountain' as seen by OSSE onboard CGRO \citep{Purcell1997_511}.
The OSSE image was reconstructed by using Singular Value Decomposition \citep[e.g.,][]{Johnstone1991_inverse_problems1}, and showed a positive latitude enhancement with a 511\,keV line flux of $(5.4 \pm 1.5) \times 10^{-4}\,\mathrm{ph\,cm^{-2}\,s^{-1}}$ in a region roughly defined by a 2D Gaussian centred at $(\ell,b) = (-1.1^\circ \pm 2.0^\circ,9.0^\circ \pm 1.3^\circ)$ with a FWHM of $11.4^\circ \pm 2.8^\circ$.
\citet{Purcell1997_511} also used the Maximum Entropy method \citep{Narayan1986_MaxEnt} which resulted in a similar feature.
The feature has been discussed in terms of supernova positron production as well as jet-activity of Sgr A*, but has later been refuted as being a image artefact, especially because the successor mission, INTEGRAL, did not find any such chimney-like structure, nor a 511\,keV disc, in the first few years of observations \citep{Knoedlseder2005_511,Jean2006_511,Churazov2005_511}.
Only with more than 7 years of observations, \citet{Bouchet2010_511} showed that there is indeed a disc component at 511\,keV, but no positive latitude enhancement \citep[see also][]{Weidenspointner2008_511}.

Another possibility for such an emission at positive but not negative latitudes with an apparent connection to the Galactic Centre are the Jovian and Neptunian Trojans.
The OSSE measurements took place between 1991-07-12 and 1997-01-05, that is roughly a timespan of 2000 days in which the Jovian Trojans moved in total $167^\circ$ across the ecliptic.
The Neptunian Trojans moved by $12^\circ$.
Within those $5.5$\,yr, the apparent epicyclic motion of the trojans disperses their emission non-uniformly as shown in Fig.\,\ref{fig:JT_NT_OSSE}.
This leads to the effect that most of the emission appears only at positive latitudes above a certain threshold flux.
In Fig.\,\ref{fig:JT_NT_GCE}, we show the same effect for the timespan of $\sim 21$\,yr since the launch of the INTEGRAL satellite.
Also in this case, the resulting emission from the Jovian Trojans would be asymmetric even though Jupiter orbited about twice around the Sun during this period.

\begin{figure}[!t]%
	\centering
	\includegraphics[width=1.0\columnwidth]{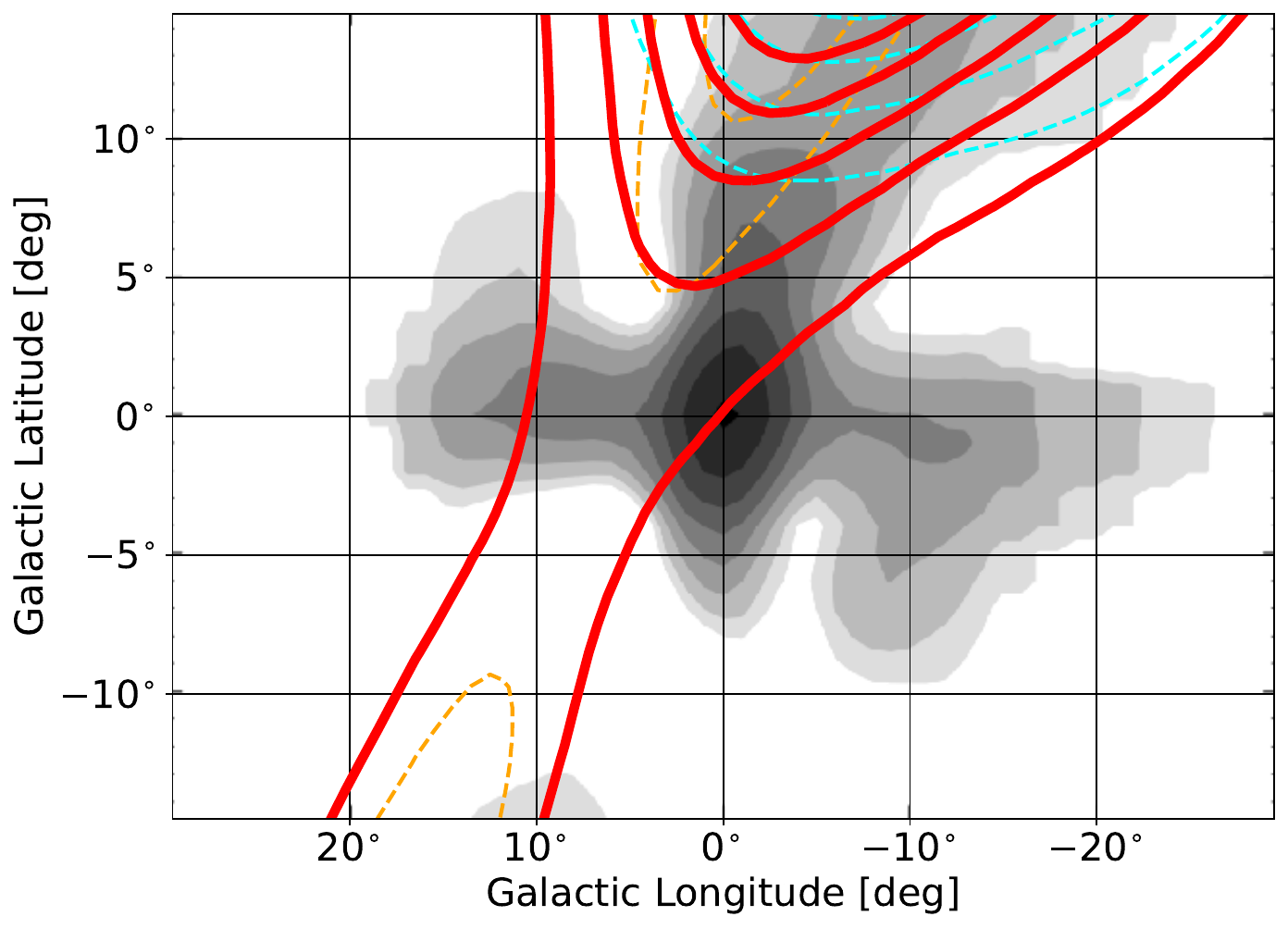}
	\caption{Image reconstruction of the 511\,keV emission from OSSE by \citet{Purcell1997_511} (gray-scale image), together with the possible cumulative contributions from the Jovian (orange; 33, 66, 80\% levels) and Neptunian (cyan; 15, 25, 35\% levels) Trojans from 1990-01-01 to 1997-12-31, and the combined model (red; 15, 25, 35\% levels). The Jovian and Neptunian Trojans clearly overlap with the `OSSE fountain', the serendipitous, and never observed afterwards, positive latitude enhancement.}
	\label{fig:osse_fountain_trojans}
\end{figure}

Assuming the Main Belt Asteroid and Kuiper Belt Object contributions to be weak, we can compare the OSSE image with the cumulative effect of the Jovian and Neptunian Trojans in the respective measurement timespan.
Fig.\,\ref{fig:osse_fountain_trojans} shows the OSSE 511\,keV map from \citet{Purcell1997_511} together with the possible contributions from the trojans.
A large overlap from their combined fluxes with the OSSE positive latitude enhancement is visible.
We do not claim that this is indeed the reason for the `OSSE 511\,keV fountain', but only illustrate one possibility here that has not been considered before.
A thorough time-dependent analysis of the ecliptic emission with more than 20\,yr of INTEGRAL/SPI measurements might reveal if this enhancement is real or not.

\subsection{Mimicking / Masking Galactic Centre features}\label{sec:GCE_mimic}
\begin{figure}[!t]%
	\centering
	\includegraphics[width=1.0\columnwidth]{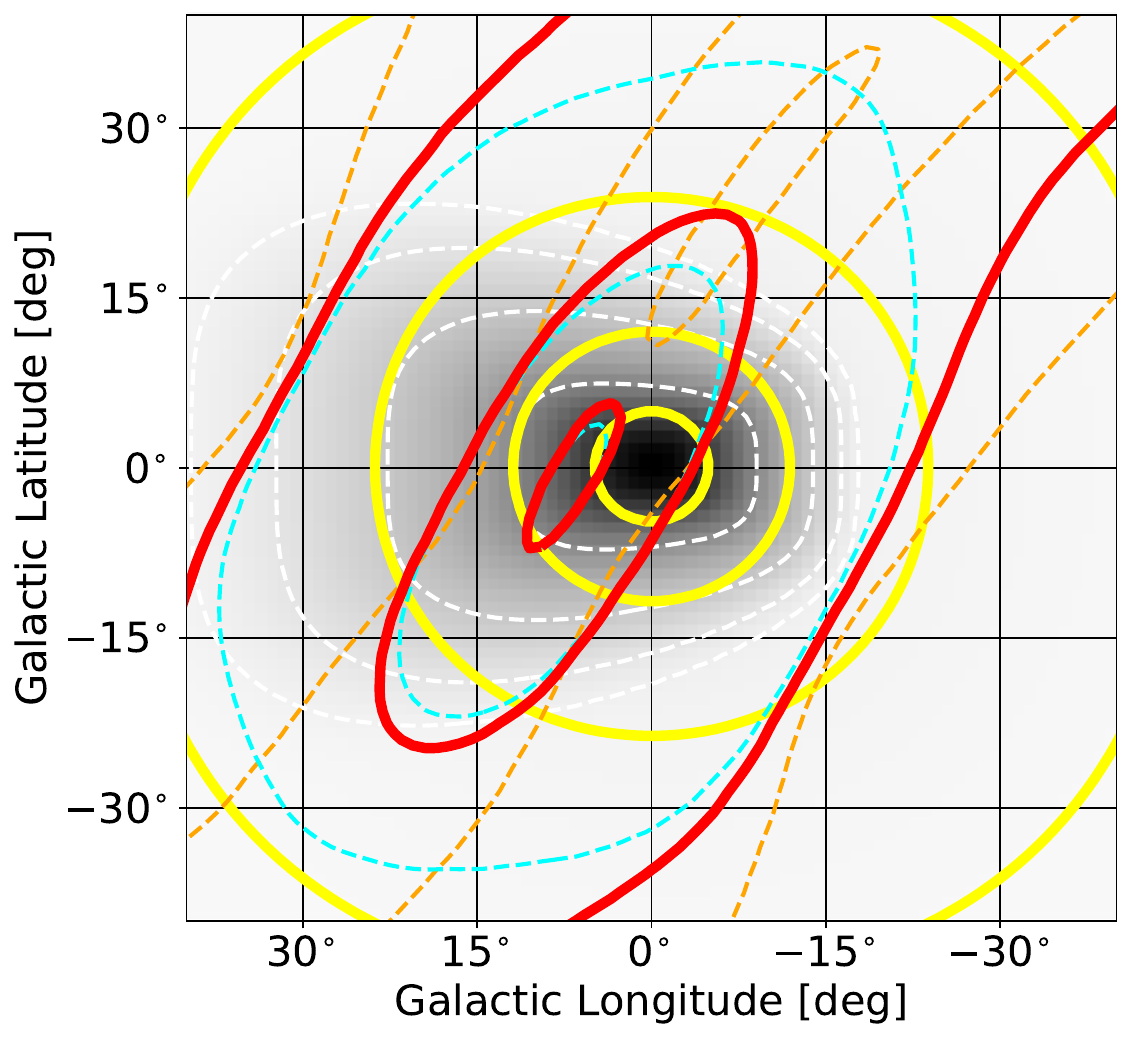}
	\caption{Image composition towards the Galactic Centre used in Fermi/LAT analyses \citep[e.g.,][gray-scale image]{Macias2018_LATGeV,Macias2019_GeVBulge,Bartels2018_GeVexcess_stars} as well as INTEGRAL/SPI analyses \citep[e.g.,][]{Siegert2022_511}, together with the possible cumulative contributions from the Jovian (orange; 5, 50, 95\% levels) and Neptunian (cyan; 5, 50, 95\% levels) Trojans from 2004-01-1 to 2024-01-01, and the combined model (red; 5, 50, 95\% levels). The extended emission towards positive longitudes coincides with the combined emission from Jovian and Neptunian Trojans, which may dilute / skew a halo-like emission component (yellow; NFW) towards the used boxy bulge \citep[white contours;][]{Freudenreich1998_BoxyBulge_COBE}. We note that the extended emission at positive longitudes would then change with time which may result in different bulge and disk sizes/extents for analyses at different times/exposures.}
	\label{fig:GCE_NFW_trojans}
\end{figure}

The so-called `Galactic Centre Excess' in GeV photons \citep[e.g.,][]{Goodenough2009_GeVDM,Macias2018_LATGeV,Macias2019_GeVBulge,Bartels2018_GeVexcess_stars} may also be related to GeV emission along the ecliptic from a varying foreground.
Excesses may originate from certain positions as a result of using a static model instead of variable one so that photons are wrongly assigned to more or less extended emission templates.
One particularly interesting case is indeed the Galactic Centre Excess because there are at least two competing models:

Either the emission is due to dark matter annihilations of WIMP-like particles into Standard Model particles and subsequent hadronisation which leads to prompt photons, or the emission is correlated with the (old) stellar population towards the direction of the bulge, including the boxy bulge by \citet{Freudenreich1998_BoxyBulge_COBE}, the nuclear stellar disk and the nuclear stellar bulge \citep{Launhardt2002_NB}.
In the latter case, the emission profile would be slightly peaked towards the Galactic Centre from the nuclear bulge and asymmetric along longitudes because of the tilted bar.
This effect is also observed to fit the 511\,keV emission in the Galaxy \citep[e.g.,][]{Siegert2022_511}.
However, the extended, apparently-larger, scale height of the bulge towards positive longitudes overlaps with the cumulative effect of the Neptunian and Jovian Trojans after the year 2000, that is, when the INTEGRAL (511\,keV, since 2002) and Fermi (GeV, since 2008) satellites started observing with higher sensitivity than previous instruments.
In Fig.\,\ref{fig:GCE_NFW_trojans}, we show the $\gamma$-ray emission towards the Galactic bulge including the boxy bulge that is typically used to characterise the old stellar population (grey-scale image), the appearance of a squared Navarro-Frenk-White \citep[NFW;][]{Navarro1997_NFW} profile (yellow contours), and the cumulative effect of the Jovian and Neptunian Trojans within a timescale of 20\,yr (red contours).
Clearly, the NFW profile and the trojans could mimic the more extended emission at positive longitudes if the time variability is not taken into account in the analysis.

We note again that these considerations are qualitative suggestions since the absolute fluxes of the asteroid populations are unknown.
We do not claim any detections and want to point out that the time variable foreground \emph{could} mimic these signatures, and therefore might be key to identifying (Galactic) dark matter signals, for example.
Future analyses of these large, decade-scale, datasets should certainly take into account that there could be a contribution of the time-variable $\gamma$-ray albedo.

In both cases, the `OSSE fountain' and the `Galactic Centre Excess', the additional emission component would not only show up near the Galactic Centre but also around coordinates $(\ell,b) = (150^\circ \pm 30^\circ, -40^\circ \pm 20^\circ)$ (see Figs.\,\ref{fig:JT_NT_OSSE} and \ref{fig:JT_NT_GCE}).
However, the emission in this region is typically not analysed as studies focus on the Galactic Centre, bulge, and disk.
A thorough analysis of the Galactic emission should therefore also take into account the ecliptic emission to fully ensure consistency of the used templates.

\section{Discussion}\label{sec:discussion}
\subsection{Detectability with current and future instruments}\label{sec:measurements}
The $\gamma$-ray emission spectrum of asteroids has first been calculated by \citet{Moskalenko2008_GRalbedoSS} based on their work of the albedo emission of the Moon in \citet{Moskalenko2007_GRalbedoMoon}.
The authors took into account objects with a radial size of 100\,cm which results in approximately the number of asteroids given in Tab.\,\ref{tab:full_sssb_models}.
Based on these earlier works, and combined with the temporal, that is, spatial, variability here, the interesting targets are either the entire ecliptic or, again, the region $120^\circ$ away from the Galactic bulge around $(\ell,b) = (150^\circ \pm 30^\circ, -40^\circ \pm 20^\circ)$ as the Jovian and Neptunian Trojans would overlap there in two of three cases every 4.5\,yr (Fig.\,\ref{fig:JT_NT_overlap}).
We advocate for full-sky analyses instead of searches for enhanced emission at specific points in time as the signals would be weak:

For example, the total 511\,keV emission from the $\gamma$-ray albedo of all components combined (without the Oort cloud, see Sec.\,\ref{sec:isotropic_emission}) would amount to $\sim 4 \times 10^{-4}\,\mathrm{ph\,cm^{-2}\,s^{-1}}$ \citep{Moskalenko2008_GRalbedoSS}, with order of magnitude variations due to the dependency on the Solar cycle.
Compared to the total 511\,keV flux of the Galaxy of $(28 \pm 3) \times 10^{-4}\,\mathrm{ph\,cm^{-2}\,s^{-1}}$ (without isotropic components) \citep{Siegert2016_511}, this would be about $14 \pm 2$\% of the total emission, and therefore easily within reach for INTEGRAL/SPI.
The reason why this component has not been `detected' so far is that it has not been analysed appropriately.
The bulge 511\,keV emission alone amount to $\sim 10 \times 10^{-4}\,\mathrm{ph\,cm^{-2}\,s^{-1}}$ of which only 2\% on average originate from the asteroid albedo \citep{Moskalenko2008_GRalbedoSS}.
The ecliptic emission is therefore even more extended than the Galactic emission which makes it difficult to detect:
For example, the Galactic disk in 511\,keV took about 8--10\,yr of observations \citep{Bouchet2010_511,Skinner2014_511,Siegert2016_511} to be unambiguously \citep{Weidenspointner2008_511} detected.
The INTEGRAL exposure time along the ecliptic is much smaller than that of the Galactic plane which prevents straight-forward detections.
A full 20\,yr all-sky 511\,keV data analysis with INTEGRAL/SPI that includes the temporal variability of the asteroids' $\gamma$-ray albedo, however, might be able to disentangle the Galactic and the ecliptic emission.

Likewise, the albedo emission spectrum from 0.1--10.0\,MeV is roughly a power-law with a slope of $-1$ \citep{Moskalenko2008_GRalbedoSS}, with several $\gamma$-ray lines from nuclear excitation showing the composition of the rocks.
At hard X-rays therefore, the diffuse emission spectrum might also extend along the ecliptic with a similar flux level as the 511\,keV line.
The low-energy spectrum might however change due to the increasing number of smaller ($< 100$\,cm) objects in which the scattering depth of cosmic rays is insufficient to produce a self-similar spectrum for all sizes.
Around 100--400\,MeV, the emission spectrum shows an exponential cutoff so that GeV emission above $\sim 10$\,GeV would be suppressed.
Nevertheless, the emission might readily be seen by more than 15\,yr of observations with Fermi/LAT if the time variability is taken into account properly.

The future Compton Spectrometer and Imager, COSI \citep{Tomsick2019_COSI,Tomsick2023_COSI}, will perform an all-sky survey, similar to Fermi, which will make it easier to identify the ecliptic emission in the MeV range because if will scan along, below, and above the ecliptic.
With its $1\pi$\,sr field of view, COSI will observe each point in the sky for 16\,Ms during its nominal 2\,yr mission, superseding the sensitivity of the pointing instrument SPI (field of view $\sim 200\,\mathrm{deg^2} \approx 0.06\,\mathrm{sr}$) across the entire sky due to its higher grasp (effective area times field of view).

\subsection{Spectral models}\label{sec:spectra}

Based on the collisional production of smaller objects by larger objects \citep{Dohnanyi1969_asteroids}, one would expect more and smaller asteroids, but which would show a different spectrum compared to the larger ones because the cosmic rays may not deposit their entire energy into the smaller objects.
This would lead to a brighter but distorted spectrum compared to \citet{Moskalenko2008_GRalbedoSS}.
While the exact spectral shape is certainly of importance in the search of residual ecliptic emission, a full calculation is beyond the scope of this work.
Furthermore, the true normalisation of the local cosmic-ray spectrum is uncertain to some degree and furthermore the true composition of the asteroids, and the distribution thereof (iron, icy, mixture), are unknown so that only individual assumptions may be calculated, similar to \citet{Moskalenko2008_GRalbedoSS}.
We leave the full discussion of the spectral models to a future study.

\subsection{Isotropic emission from the Oort Cloud}\label{sec:isotropic_emission}
The Oort Cloud, supposedly the most distant and most populated asteroid (comet) accumulation, would result in an almost isotropic hard X-ray to $\gamma$-ray flux.
The cosmic-ray spectrum at the position of the Oort Cloud would experience no influence from the Solar modulation potential so that its brightness throughout observations would be constant.
On a median level, the Oort Cloud could have a similar flux level than the other four major components, which means that its contribution may make about 25\% of the total albedo emission of the ecliptic components.

For $\gamma$-ray lines, this would create a so-far neglected isotropic component, and especially a measurable one for the 511\,keV line of about $10^{-4}\,\mathrm{ph\,cm^{-2}\,s^{-1}}$ or $\sim 10^{-5}\,\mathrm{ph\,cm^{-2}\,s^{-1}\,sr^{-1}}$.
Currently, soft $\gamma$-ray telescopes are blind to isotropic emission if they use a coded aperture mask \citep{Siegert2022_gammaraytelescopes}.
Future MeV telescopes that utilise Compton scattering to detect photons have fewer problems to observe isotropic emission.
A similar discussion for isotropic GeV emission can be found in \citet{Moskalenko2009_CGB_gammaray_albedo}.

In terms of determining the Local Interstellar Cosmic-ray spectrum \citep{Vos2015_LICRS}, such an isotropic $\gamma$-ray line emission would make an excellent tracer:
While currently, only the Voyager probes can measure the (low-energy) cosmic-ray spectrum outside the sphere of influence of the Sun \citep{Stone2013_Voyager1_CR}, soft $\gamma$-ray telescopes could measure the flux of individual lines, such as \nuc{Fe}{56} (0.85, 1.24\,keV), \nuc{O}{16} (6.16\,MeV), and \nuc{C}{12} (4.44\,MeV), and therefore directly constrain the shape \emph{and} amplitude of cosmic rays.
The fluxes of these lines for asteroid populations have not been extensively calculated in the literature and we leave this to a future study.
In fact, the same measurement of nuclear de-excitation lines in supernova remnants, for example, would lead to a direct measurement of the low-energy cosmic-ray spectrum in these objects \citep{Summa2011_CasA,Benhabiles-Mezhoud2013_DeExcitation}, and measurements of their line profiles would reveal accretion flows around black holes, for example \citep{Yoneda2023_nuclear_lines}.

\subsection{Reduction of Galactic fluxes}\label{sec:galactic_fluxes}
The overlap of the ecliptic emission, expected from cosmic-ray interactions with asteroids, with the Galactic bulge and disk will enhance the total flux level when only limited regions of interest are used for $\gamma$-ray data analyses.
If the albedo emission is not taken into account with its time dependence, the additional flux towards the Galactic bulge region is about 1--10\%, depending on energy.
As a function of time, the additional flux may decrease to 0.1\% at times for hard X-rays ($\lesssim 100$\,keV), but can be enhanced to about 20\% in the few 100\,MeV range.
This means, the Galactic bulge $\gamma$-ray flux from $10^{4}$--$10^{10}$\,eV and nearby emission components, such as the disk within $|\ell| \lesssim 20^\circ$ and the halo around $\ell \approx 0^\circ$ within $|b| \lesssim 30^\circ$, that is, the size of the Fermi Bubbles \citep[e.g.,][]{Su2010_fermibubbles}, are actually \emph{weaker}.
This will have a large impact on the interpretations of several unsolved problems, such as the `Positron Puzzle' \citep{Prantzos2011_511,Siegert2023_511}, the `Galactic Centre Excess' \citep[e.g.,][]{Goodenough2009_GeVDM,Macias2019_GeVBulge,Bartels2018_GeVexcess_stars}, as well as the cosmic-ray flux in the Galactic Centre \citep[e.g.,][]{HESS2016_pevatron} and the consequences for low-energy cosmic-rays \citep[e.g.,][]{Siegert2022_diffuseemission}, the supposed dark matter annihilation flux normalisation \citep[e.g.,][]{Karwin2017_DM_MW,Berteaud2022_SPI_PBH}, nucleosynthesis mass estimates \citep[e.g.,][]{Diehl2021_NucSyn} and the subsequent derived parameters, such as the star formation rate, the supernova rate \citep[e.g.,][]{Siegert2023_PSYCO} or the novae rate \citep[e.g.,][]{Shafter2017_novarate,Siegert2021_BHMnovae}, among others.
While the effect of the reduced fluxes may be small compared to the Galactic emission, the effects of all these individual problems together may be alleviated by simply taking into account the time-variable diffuse $\gamma$-ray foreground emission of the Solar System.

\subsection{Future work}\label{sec:future_work}
As discussed above, the $\gamma$-ray albedo emission has not been detected so far, mainly because it was not searched for, except in \citet{Moskalenko2008_GRalbedoSS} with CGRO/EGRET.
A clear signal may only be found if the time variability is actually taken into account in long-exposure, full-sky datasets across multiple energies.
A next step in terms of data analysis would therefore be the complete re-analysis of current MeV and GeV datasets that take into account the albedo emission.

Especially for the forward modelling of the emission templates when using the instrument response functions, a more accurate spectrum might be needed.
In particular for the low-energy part of the emission spectrum below 10\,MeV, the smaller asteroids below the 100\,cm radius scale may be of importance as they would shape the spectrum with a low-energy cutoff that depends probably on the composition of the rocks.
Also the nuclear de-excitation $\gamma$-ray lines from the albedo emission would show the average composition of the asteroids so that more detailed simulations of the interactions of cosmic rays with asteroids of different sizes, and potentially shapes, should be performed.

In fact, the cosmic-ray impact on the asteroids is not the only emission component that would contribute to their albedo.
Similar to the works by \citet{Moskalenko2007_GRalbedoMoon} and \citet{Churazov2008_EarthMoonSunAlbedo} in which the Moon, Earth, and Sun $\gamma$-ray albedo has been calculated by including the Cosmic Gamma-ray Background \citep[e.g.][]{Ajello2009_CGB,Inoue2013_CGB}, also the asteroids should reflect some portion of this isotropic $\gamma$-ray component.
This would make it possible even for coded-mask telescopes to measure the MeV Cosmic Gamma-ray Background as the isotropic emission would be concentrated towards smaller regions.
The Cosmic Gamma-ray Background reflection on asteroids and their cumulative effect as seen from Earth has not been calculated in the literature.

\section{Conclusion}\label{sec:conclusion}
In this work we calculated the appearance of the cosmic-ray induced $\gamma$-ray albedo of asteroid accumulations in the Solar System.
Based on the relative motion of the asteroids with respect to Earth, the resulting emission features move in epicycles along the ecliptic which may mimic extended emission unless properly taken into account in $\gamma$-ray data analyses.
The Solar modulation potential enhances the time variability even further as asteroid accumulations closer to the Sun, such as the Main Belt Asteroids (2.8\,AU) or the Jovian Trojans (5.0\,AU), will be shielded more or less from incoming cosmic rays than outer asteroids such as the Neptunian Trojans (28\,AU), Kuiper Belt Objects (40\,AU) or the hypothetical Oort Cloud ($>4000$\,AU).

We show that the time variable foreground can potentially explain serendipitous emission features such as the `511\,keV OSSE fountain' or enhance the case for a spherical `Galactic Centre GeV Excess' as the ecliptic albedo intersects with the Galactic bulge roughly round $0^\circ \lesssim \ell \lesssim 10^\circ$, inclined by $\sim 62^\circ$.
This conjecture of enhanced MeV--GeV $\gamma$-ray emission due to the asteroid cosmic-ray albedo is further corroborated by the fact that the Jovian and Neptunian Trojans appear exactly at the right positions at exactly the right times of observations by the respective instruments (OSSE: 1991--1997; INTEGRAL: 2002--now; Fermi: 2008--now).
We do not claim that the asteroid albedo does explain these special emission features but rather that, on a qualitative basis, the overlaps of the Galactic emission and cumulative albedo emission are remarkably consistent.
Proving these conjectures would require full-sky full-mission dataset analyses at both MeV and GeV energies.

The non-detections of the Solar System $\gamma$-ray albedo are, so far, not surprising because there have been no dedicated searches with proper emission templates, and the foreground is certainly not static but changes as a function of time, energy, and the Solar cycle.
The relative amplitudes of the components are unknown and uncertain between one and three orders of magnitude; 
each component, be it the Main Belt Asteroids, Jovian Trojans, Neptunian Trojans, Kuiper Belt Objects, or the Oort Cloud, could be the dominant one, even if further away from Earth.
INTEGRAL/SPI and Fermi/LAT already have the required sensitivity to detect the asteroid $\gamma$-ray albedo if modelled consistently as time-variable across their 22 and 15\,yr of observations, respectively, in full-sky datasets.
Future MeV telescopes will be even better suited to search for this emission because of a more uniform exposure and larger field of view in the case of COSI, for example, seeing the entire sky within one day.
In order to properly model the $\gamma$-ray albedo, more work is required to calculate the emission spectrum of asteroids below the 100\,cm radius scale as they would provide a low-energy cutoff because the cosmic rays cannot deposit their entire energy in such small but much more numerous objects.
In addition will the asteroids reflect the Cosmic Gamma-ray Background which would make it possible also for coded-mask telescopes, such as INTEGRAL/SPI, to measure the isotropic emission beyond the currently reliable estimates up to 400\,keV.

As has already been discussed in a previous study by \citet{Moskalenko2008_GRalbedoSS}, the total Galactic flux from $10^4$--$10^{10}$\,eV is \emph{weaker} by 0.1--20\%, depending on time and energy, than reported in previous studies.
This has a large impact on the parameters derived for cosmic-ray spectra and amplitudes, limits on dark matter annihilation cross sections, as well as nucleosynthesis yields and supernova rates, for example.
Most of the MeV--GeV $\gamma$-ray measurements that include the Galactic bulge would result in a smaller flux when taking the time-variable diffuse $\gamma$-ray foreground into account, so that also the derived parameters would obtain different values, with potentially different -- but more reliable -- statistical and smaller systematic uncertainties.


\bibliographystyle{aa} 
\bibliography{thomas} 


%


\end{document}